\documentclass[fleqn,usenatbib,useAMS]{mnras}

\usepackage{graphicx}	
\usepackage{amsmath}	
\usepackage{longtable}
\usepackage{booktabs}
\usepackage{rotating}
\usepackage{lscape}
\usepackage{array}
\usepackage{caption}

\usepackage[T1]{fontenc}
\usepackage{ae,aecompl}
\usepackage{color}

\usepackage{newtxtext,newtxmath}

\title[The FP for HBLs]{The Fundamental Planes of Black Hole Activity for High-Synchrotron-Peaked BL Lacertae Objects}

\author[Q. C. Long et al.]{
Qing-Chen. Long,$^{1,2}$
Ai-Jun. Dong$^{1,2}$\thanks{E-mail: aijdong@gznu.edu.cn}
and Qi-Jun. Zhi$^{3}$
\\
$^{1}$School of Physics and Electronic Science, Guizhou Normal University, Guiyang 550025, People’s Republic of China; aijdong@gznu.edu.cn\\
$^{2}$Guizhou Provincial Key Laboratory of Radio Astronomy and Data Processing, Guizhou Normal University, Guiyang 550025, People’s Republic of China\\
$^{3}$College of Physics, Guizhou University, 550025 Guiyang, People’s Republic of China\\
}

\date{Accepted XXX. Received YYY; in original form ZZZ}

\pubyear{\the\year{}}

\begin{document}
\label{firstpage}
\pagerange{\pageref{firstpage}--\pageref{lastpage}}
\maketitle

\begin{abstract}
The radio--X-ray correlation and Fundamental Plane (FP) of black hole activity can serve as a diagnostic tool for the origin of X-ray emissions. There was a scaling relation between radio and X-ray emissions for High-synchrotron-peaked BL Lacertae objects (HBLs), i.e., $L_{\rm{R}}\propto L_{\rm{X}}^{0.64}$, which can be explained by ADAF-dominated mode or synchrotron cooling (Syn-c). However, many results of studying blazar physics show that the X-ray emissions of HBLs are mainly produced by the synchrotron process of jets. Therefore, Syn-c appears to provide a plausible explanation for this relation. To further clarify the origin of X-ray emissions of HBLs, we constructed a sample containing 69 HBLs in this paper to re-investigate their radio--X-ray correlation and FP. Considering the Doppler beaming effect, we find that the intrinsic radio--X-ray correlation and FP of HBLs are $L_{\rm R,int}\propto L_{\rm X,int}^{0.68}$ and $\log L_{\rm R,int}=(0.57\pm0.06)\log L_{\rm X,int}+(0.33\pm0.11)\log M_{\rm BH}+(12.65\pm2.00)$, respectively. Our results agree with the scaling relation mention above, which suggests these scaling relations are not artificial. By employing the theoretical model of Syn-c, we find that these shallow radio--X-ray correlations and FP are caused by Syn-c, which implies that the X-ray emissions of HBLs may be produced by rapidly cooling, high-energy electrons accelerated at a shock. This is consistent with results of the recent X-ray polarization observations of HBLs. Our results provide the observational evidence of $L_{\rm R}\propto L_{\rm X_{\text{Syn-c}}}^{0.6\sim0.7}$.
\end{abstract}




\section{Introduction}\label{Intro}

Active galactic nuclei (AGNs) reside at the centers of nearly all massive galaxies. They emit enormous amounts of energy across the entire electromagnetic spectrum that usually outshine all stars in their host galaxies. This phenomenon arises from the accretion onto supermassive black holes (SMBH: $M_{\rm BH}\sim10^{6-10}\,M_{\odot}$). AGNs can be classified into radio-loud AGNs (RL-AGNs) and radio-quiet AGNs (RQ-AGNs) based on their radio-loudness parameter, $R\equiv F_{\rm{5GHz}}/F_{\rm{4400\text{\AA}}}$, where $F_{\rm{5GHz}}$ and $F_{\rm{4400\text{\AA}}}$ represent the flux density at rest-frame radio 5\,GHz and  optical B-band $\rm{4400\text{\AA}}$ \citep[e.g.,][]{Kellermann1989}. Only $10\%-20\%$ of the AGNs belong to RL-AGNs ($R>10$), while the remaining ones are RQ-AGNs ($R<10$)  \citep[e.g.,][]{Ivezic2002optical,Kellermann2016radio}. Typically, RL-AGNs have powerful relativistic jets that are absent in RQ-AGNs \citep{Padovani2017}. Consequently, the emitting properties of RL-AGNs are frequently considered to be closely related to the physics underlying relativistic jets.

Blazars are typical RL-AGNs with a relativistic jet towards the observer \citep{Urry1995}. This jet orientation leads to the boosting in their broad-band emissions and confers them with extreme properties \citep[e.g., rapid variability, high polarization, and high luminosity.][]{Wills1992survey,Lichti2008integral,Raiteri2017blazar,Zhao2024energy}. Based on the equivalent width (EW) of the optical emission lines, blazars are generally classified into flat-spectrum radio quasars (FSRQs) and BL Lacertae objects (BL Lacs). FSRQs display strong broad emission lines with $\rm{EW}>5\text{\AA}$, while BL Lacs have no or weak broad emission lines characterized by $\rm{EW}<5\text{\AA}$ \citep[e.g.,][]{Scarpa1997high}. The weak/lack of broad emission lines may suggest the existence of inefficient advection-dominated accretion flows (i.e., ADAF), resulting in insufficient energy to photoionize the clouds of the broad line region \citep[e.g.,][]{Ghisellini2011transition,Sbarrato2014jet,Chen.Y-Y2023jet}. It is noted that BL Lacs exhibit a relativistic jet aligned with our line of sight. This geometric alignment will give rise to relativistic beaming effects, quantified by the Doppler factor $\delta$, which can amplify the observed fluxes in the X-ray and radio band \citep{Urry1995, Wu2007,Nieppola2008A&A,Yang2022beaming}. Therefore, when performing the physical inference for blazars through multi-wavelength diagnostics, it is necessary to utilize Doppler corrected intrinsic luminosity, $L_{\rm{\nu,int}}\approx L_{\rm{\nu,obs}} \cdot \delta^{-(2+\alpha_{\nu})}$, where $\alpha_{\nu}$ is the spectral index for the corresponding observational band \citep[e.g.,][]{Long.Q-C2025FP}. 

The spectral energy distribution (SED) of the BL Lacs shows a double hump structure \citep{Fossati1998unifying,Donato2001hard,Wu2007,Ghisellini2011transition,Fan2016spectral,Yang2022spectral,Ajello2022fourth}. The lower-energy hump, characterized by synchrotron radiation, spans from the radio to the X-ray bands. In contrast, the higher-energy hump lies in the range of MeV and TeV energy bands and is characterized by inverse Compton (IC) scattering of relativistic electrons \citep{Fossati1998unifying,Donato2001hard,Wu2007,Zhao2024energy}. Based on the location of synchrotron peak frequency ($\rm \nu_p^{syn}$) in the SED, BL Lacs can be classified into low-synchrotron-peaked BL Lacs (LBL, $\rm log(\nu_p^{syn}/Hz)<14$), intermediate-synchrotron-peaked BL Lacs (IBL, $\rm 14<log(\nu_p^{syn}/Hz)<15.3$) and high-synchrotron-peaked BL Lacs \citep[HBL, $\rm log(\nu_p^{syn}/Hz)>15.3$, see][]{Fan2016spectral}. FSRQs are usually the powerful sources with $\rm log(\nu_p^{syn}/Hz)\lesssim14$. The different energy distributions might imply different accretion physics \citep{Wang.J-M2002}. \cite{Keenan2021relativistic} suggested that FSRQs, LBLs, and IBLs are strong jet sources that display the radiatively efficient accretion mode, while HBLs are weak jet sources associated with the radiatively inefficient accretion mode. \cite{Chen.Y-Y2023jet} found that 62\% of their HBL sample have pure optically thin ADAF, while 94\% of HBLs can be explained by the hybrid mode of ADAF+standard thin disk \citep[][SSD]{Shakura1973}. \cite{Zhao2024energy} showed 96.8\% of their HBLs exhibit pure optically thin ADAF, and 2.1\% of HBLs belong to ADAF + SSD mode.

The radio--X-ray correlation ($L_{\rm{R}}\propto L_{\rm{X}}^{\xi_{\rm{RX}}}$) and the fundamental plane of black hole activity (FP, $L_{\rm{R}}=\xi_{\rm{RX}}\log L_{\rm{X}}+\xi_{\rm{RM}}\log M_{\rm{BH}}+c$) can serve as powerful diagnostics to constrain the accretion mode and trace the origin of X-ray emissions in black hole (BH) sources \citep{Merloni2003,yuan2005radio,Kording06RefiningA&A,Plotkin12,dong2014new,Wang2024FP,Long.Q-C2025FP}. In theoretical model, the relativistic jets are produced from the innermost regions of the accretion disk, so the jet variables should depend on two fundamental parameters that determine the conditions in the inner accretion disc, namely, dimensionless accretion rate ($\dot{m}=\dot{M}/\dot{M}_{\rm  Edd}$, $\dot{M}_{\rm Edd}$ is the Eddington accretion rate) and BH mass \citep[$M_{\rm BH}$, see][]{Heinz2003}. It is widely believed that the radio emissions are produced by the synchrotron self-absorption of jets. Using the standard formulae of synchrotron emission, \cite{Heinz2003} showed that the radio emissions depend non-linearly on $M_{\rm BH}$ and $\dot{m}$: $L_{\rm R}\propto(M_{\rm BH}\dot{m})^{17/12}\propto \dot{M}^{1.42}$. In the BH system, there are several origins of X-ray emissions, the prevailing origins include ADAF-dominated mode, SSD+corona mode, canonical synchrotron radiation (C-syn) from jets, synchrotron cooling (Syn-c) from jets, and IC process from jets \citep[see][]{Merloni2003,Heinz2003,Heinz2004,yuan2005radio,Plotkin12,Fan2016spectral}. The observed X-ray emissions are likely a superposition of several components, it is just that a specific component mainly dominates X-ray emissions under the certain conditions \citep{Plotkin12}.

In ADAF-dominated mode, the X-ray emissions follow the scaling relation as $L_{\rm X_{\rm ADAF}}\propto\dot{M}^{2\sim2.3}$ \citep[e.g.,][]{Merloni2003,yuan2005radio}, yielding $L_{\rm R}\propto L_{\rm X_{\rm ADAF}}^{0.6\sim0.7}$. This correlation has been verified by the observed results in many previous works \citep[e.g.,][]{Merloni2003,Li2008black,dong2015revisit,li2018black,Bariuan2022fundamental,Wang2024FP}. In the SSD+corona mode, the X-ray emissions follow the scaling relation as $L_{\rm X_{SSD}}\propto\dot{M}$ \citep{Merloni2003}. This leads to the derivation of $L_{\rm R}\propto L_{\rm X_{SSD}}^{1.42}$, which has been verified by the observed results in \cite{dong2014new}. In the scenario of C-syn from jets, the X-ray emissions follow the scaling relation as $L_{\rm X_{\text{C-syn}}}\propto\dot{M}^{1.1\sim1.25}$ \citep[e.g.,][]{Merloni2003,yuan2005radio}, from which one can derive $L_{\rm R}\propto L_{\rm X_{\text{C-syn}}}^{1.1\sim1.29}$, which has been verified by the observed results of RL-AGNs \citep[e.g.,][]{Wang2006black,yuan2009revisiting,Xie2017fundamental,Liao2020x,Bariuan2022fundamental,Wang2024FP,Long.Q-C2025FP}. For the IC origin from jets, $L_{\rm X_{IC}}\propto(U_{e}/U_B)L_{\rm X_{\text{C-syn}}}\approx L_{\rm X_{\text{C-syn}}}$ \citep[$U_e/U_B\sim1$, e.g.,][where $U_e$ and $U_B$ are the electron and magnetic field energy density, respectively]{Majumdar2025}. Hence, we predict $L_{\rm X_{IC}}$ roughly agrees with the scaling relation of $L_{\rm X_{\text{C-syn}}}$, which can also be found in the observed results from \cite{Long.Q-C2025FP}. It is clear that the $L_{\rm R}\propto L_{\rm X}^{1.42/\xi_{\rm X}}$ relation and FP are regulated by the origin of X-ray emissions \citep[also see][]{Long.Q-C2025FP}. However, \cite{Merloni2003} pointed out that the synchrotron cooling origin of X-ray emissions may also lead to the scaling relation as $L_{\rm R}\propto L_{\rm X}^{0.6\sim0.7}$, which has been further verified by \cite{Heinz2004} using the theoretical model of synchrotron cooling, indicating this shallow correlation is not unique to ADAF origin. However, the relationship $L_{\rm R} \propto L_{\rm X_{\text{Syn-c}}}^{0.6\sim0.7}$ remains a theoretical prediction without observational evidence so far. This absence will be addressed in this paper.

\cite{Donato2005six} found that the strong jet blazars (FSRQs and LBLs) exhibit a steeper radio-X-ray correlation ($L_{\rm{R}}\propto L_{\rm{X}}^{1.06}$), whereas HBLs follow a shallower relation ($L_{\rm{R}}\propto L_{\rm{X}}^{0.64}$). It is clear that the former agrees with C-syn model, suggesting that the X-ray emissions of strong jet blazars are mainly produced by canonical synchrotron radiation from jets \citep[see also][]{Long.Q-C2025FP}. However, the latter has two possible scenarios, namely, ADAD-dominated mode and synchrotron cooling. However, the SED of HBLs is characterized by peaking synchrotron emission in the ultraviolet--soft X-ray band \citep{Fossati1998unifying,Donato2001hard}, indicating that the X-ray emissions of HBLs may still be dominated by non-thermal processes \citep[e.g.,][]{Wu2007,Fan2016spectral,Yang2022spectral}. Recent X-ray polarimetric measurements of HBLs also indicated that X-ray emissions from HBLs can be interpreted as originating from synchrotron radiation by relativistic electrons cooling in the magnetic field of jets \citep{Di-Gesu2023,Errando2024,Pacciani2025}.

To the best of my knowledge, there has never been a claim that thermal accretion flows dominate the X-ray emissions of HBLs given blazar SED and several hundred papers of studying blazar physics. Conservatively, we still believe in this paper that the X-ray emissions of HBLs are mainly emitted from the synchrotron process in jet. Therefore, we investigated in this paper whether synchrotron cooling is responsible for the $L_{\rm R}\propto L_{\rm X}^{0.64}$ relation reported by \cite{Donato2005six} for HBLs. In order to farther clarify the origin of X-ray emissions for HBLs, we compiled a larger sample containing 69 HBLs to re-explore the radio--X-ray correlation and FP for HBLs. 
The structure of the paper is organized as follows. In section\,\ref{Sample}, we describe our HBL sample in detail. In section\,\ref{METHODS AND RESULTS}, we introduce the fitting method and present our best fitting results. The results are discussed and summarized in section\,\ref{Discussion} and section\,\ref{SUMMARY}, respectively. For this paper, the cosmological parameters are selected as $H_{0} = 70\rm\,km\,s^{-1}\,Mpc^{-1}$, $\Omega_{\Lambda} =0.73$, and $\Omega_{\rm{M}} = 0.27$ \citep{dong2015revisit}.

\begin{figure*}
    \centering
    \includegraphics[width=\textwidth]{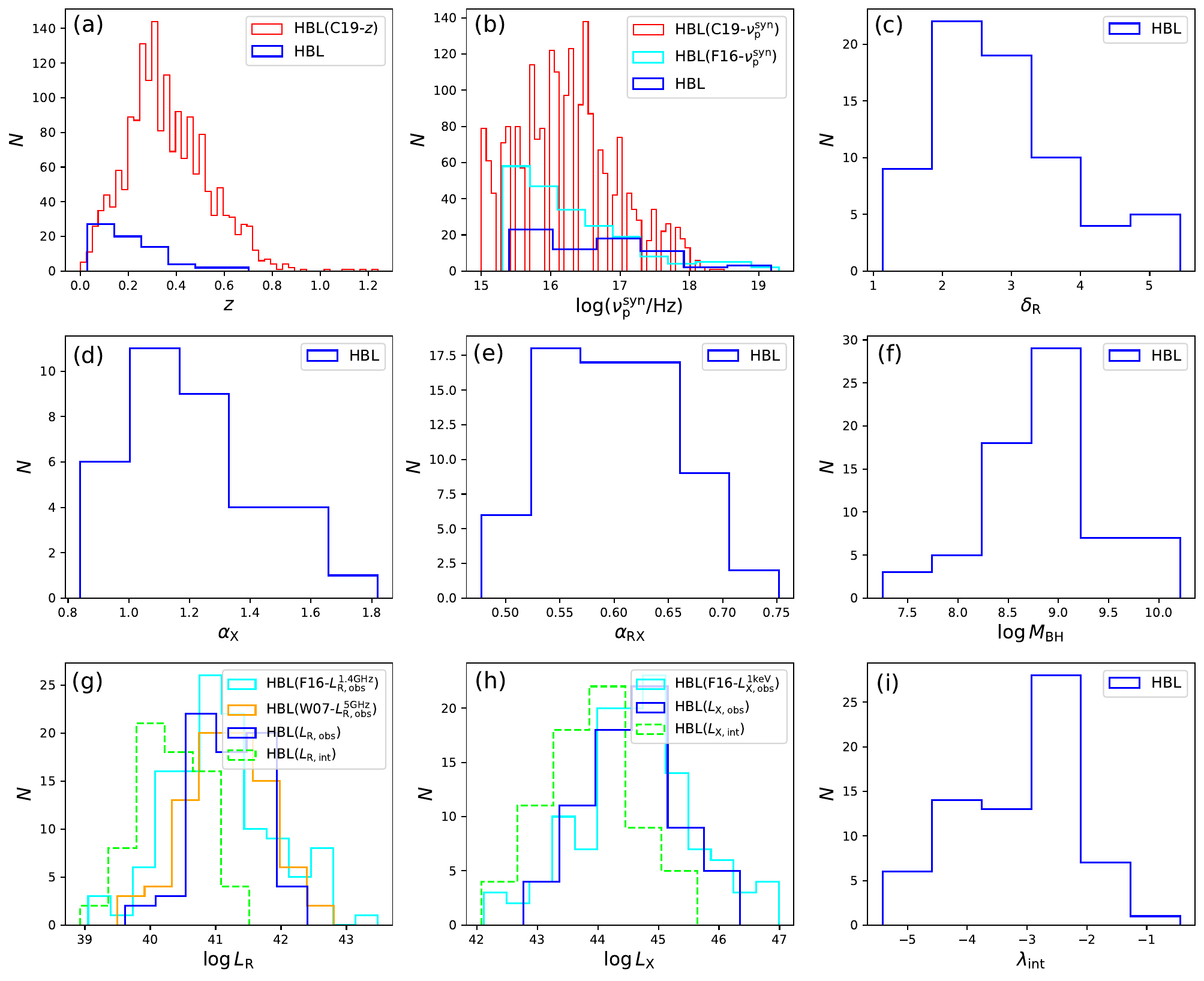}
    \caption{Distributions of the physical parameters for our HBLs and a comparison of some properties with a larger HBL sample from \citet{Wu2007}, \citet{Fan2016spectral}, and \citet{Chang2019}. The blue boxes are the distribution of observational properties for our HBLs. (a)\,Redshift: $z$, the red box is the distribution of redshift of 3HSP sample from \citet{Chang2019}. (b)\,Logarithm of the synchrotron peak frequency: $\rm log(\nu_p^{syn}/Hz)$, the red box is the distribution of $\rm log(\nu_p^{syn}/Hz)$ of 3HSP sample from \citet{Chang2019}, the cyan box is the distribution of $\rm log(\nu_p^{syn}/Hz)$ of HBL sample from \citet{Fan2016spectral}. (c)\,The 5\,GHz Doppler factor: $\delta_{\rm{R}}$. (d)\,The X-ray spectral index of HBLs with available $\Gamma_{\rm X}$: $\alpha_{\rm X}=\Gamma_{\rm X}-1$. (e)\,The broad-band spectral index from 5\,GHz radio band to 2--10\,keV X-ray band: $\alpha_{\rm RX}$. (f)\,Logarithm of the dynamic BH mass: log$M_{\rm{BH}}$. (g)\,Logarithm of radio luminosity: $\log L_{\rm R}$, the cyan box is the distribution of observational 1.4\,GHz total radio luminosity ($L_{\rm R,obs}^{\rm 1.4GHz}$) of HBL sample from \citet{Fan2016spectral}, the orange box is the distribution of observational 5\,GHz core radio luminosity ($L_{\rm R,obs}^{\rm 5GHz}$) of HBL sample from \citet{Wu2007}, the lime box is the distribution of intrinsic 5\,GHz core radio luminosity for our HBLs. (h)\,Logarithm of X-ray luminosity: $\log L_{\rm X}$, the cyan box is the distribution of observational 1\,keV X-ray luminosity ($L_{\rm X,obs}^{\rm 1keV}$) of HBL sample from \citet{Fan2016spectral}, the lime box is the distribution of intrinsic 2--10\,keV X-ray luminosity for our HBLs. (i)\,Eddington-ratio: $\lambda_{\rm{int}}=\log (L_{\rm{X,int}}/L_{\rm{Edd}})$.}
    \label{HIST}
\end{figure*}

\section{Sample}\label{Sample}
To avoid the suspicion and selection effect, it is essential to further clarify the selection criteria for our HBL sample. As mention in section\,\ref{Intro}, FSRQs+LBLs follow the scaling relation as $L_{\rm R}\propto L_{\rm X}^{1.06}$, while HBLs follow the scaling relation as $L_{\rm R}\propto L_{\rm X}^{0.64}$. In addition, \cite{Long.Q-C2025FP} collected a large sample containing 50 FSRQ, 51 LBLs, and 18 IBLs to study the radio--X-ray correlation and FP. By considering the Doppler beaming effect, they found a tightly intrinsic radio--X-ray correlation $L_{\rm R,int} \propto L_{\rm X,int}^{1.04}$ with a small scatter $\sigma_{\rm int}=0.32\,\rm dex$. This dichotomy of $L_{\rm R}\propto L_{\rm X}^{1.42/\xi_{\rm X}}$ relation between FSRQs+LBLs+IBLs and HBLs implies that their underlying physics are different. Moreover, in the theoretical model of synchrotron cooling, spectrum parameters should meet $p>3$ and $\alpha_{\rm X}>1$ \citep[see][where the $p$  and $\alpha_{\rm X}$ will be introduced in section\,\ref{New-FP}]{Merloni2003,Heinz2004}. Among blazars, HBLs seem to be the only class that fulfills this criterion \citep{Kino2002ApJ,Kino&Takahara2004,Donato2001hard,Donato2005six}. Therefore, we restricted our sample to HBLs and excluded other blazar classes.

In this work, we aim to study the radio--X-ray correlation and FP for HBLs. Following the previous work, we adopt the 5\,GHz core radio luminosity and 2--10\,keV X-ray luminosity. In addition, the Doppler-corrected luminosity should be used due to the Doppler beaming effect. Moreover, the BL Lacs should have available $\rm \nu_p^{syn}$ to ensure they are HBLs. Therefore, when collecting HBLs, they should have available 5\,GHz core radio flux, X-ray flux, BH mass, Doppler factor, and $\rm \nu_p^{syn}$. In the past, there were two meaningful works that have estimated the 5\,GHz Doppler factor for BL Lacs \citep[e.g.,][]{Wu2014some,Ye2021unification}, and there were two recent works that have provided $\rm \nu_p^{syn}$ values for BL Lacs \citep[e.g.,][]{Chang2019,Ajello2022fourth}. To obtain as large a HBL sample as possible under our selection criteria, we cross-referenced the BL Lac catalog from \cite{Wu2014some} and \cite{Ye2021unification} with the BL Lac catalog from \cite{Ajello2022fourth} and \cite{Chang2019}. In total, we have constructed a sample containing 69 HBLs with the available core radio flux, X-ray flux, BH mass, $\rm \nu_p^{syn}$, and $\delta$ at 5\,GHz ($\delta_{\rm{R}}$) (see Table \,\ref{table2}). In our sample, there are 13 HBLs (see HBLs with the Doppler factor marked with '*' in Table \ref{table2}) that are not presented in \cite{Wu2014some} and \cite{Ye2021unification}. Their Doppler factors are calculated with the Equation.\,6 in \cite{Ye2021unification} and the 5\,GHz core radio flux density ($F_{\rm{c,5GHz}}$) taken from NASA/IPAC Extragalactic Database (NED\footnote{https://ned.ipac.caltech.edu/}) or literature \citep{Yuan2018,Giroletti2004sample}. 

\subsection{The Radio Luminosity ($L_{\rm{R}}$)}
To mitigate the contaminations from the star formation and the extended emissions, the core radio luminosities at 5\,GHz are adopted in this work. In addition to 13 HBLs mentioned above, the 5\,GHz core radio flux density and $\delta_{\rm{R}}$ for rest of HBLs are taken from \cite{Wu2014some} and \cite{Ye2021unification}. Note that if the $\delta_{\rm{R}}$ was taken from \cite{Ye2021unification}, we adopted the $\delta_{\rm{R}}$ values that are estimated by using their Eq\,6 with $q=3+\alpha\,(\alpha=0)$. To explore the intrinsic physics of HBLs, we adopt the intrinsic core radio flux $F_{\nu,\rm{int}}$, where the Doppler boosting effect is considered and corrected by equation $F_{\nu,\rm{int}}=F_{\nu,\rm{obs}}\delta_{\nu}^{-q}$ (see \S\,\ref{Beaming}). 

\subsection{The X-ray Luminosity ($L_{\rm{X}}$)}
The observational X-ray fluxes of HBLs are gathered from the NED and existing literature. X-ray observations are detected by the \textit{Chandra} and \textit{XMM–Newton} are used preferentially to maintain a high precision \citep{li2018black}. For the HBLs without observations from \textit{Chandra} and \textit{XMM–Newton}, the observational X-ray fluxes from the observations of other telescopes (e.g., \textit{Swift}, \textit{BeppoSAX}) are used. If there are multiple values of observed fluxes for HBLs, the average value is adopted. In this work, the 2--10 keV X-ray luminosity is adopted. To avoid errors caused by converting through another approximate waveband, we prefer 2--10\,keV X-ray flux ($F_{\rm{2-10\,keV}}$). For the HBLs without the available $F_{\rm{2-10\,keV}}$, the X-ray flux at other bands (e.g., 0.5--7\,keV, 0.3--8\,keV, 0.3--10\,keV, 0.2--12\,keV) are extrapolated to $F_{\rm{2-10\,keV}}$ using the power law photon index $\Gamma_{\rm{X}}$ ($F_{\nu}\propto\nu^{1-\Gamma_{\rm{X}}}$):
\begin{equation}
    F_{\rm{2-10\,keV}}=F_{\rm{a-b\,keV}}\frac{\int_2^{10}\nu^{1-\Gamma_{\rm{X}}}d\nu}{\int_a^b\nu^{1-\Gamma_{\rm{X}}}d\nu}
    \label{Eq1}
\end{equation}
To reduce the errors, the $\Gamma_{\rm{X}}$ and the X-ray flux densities are taken from the same literature. For HBLs without the available $\Gamma_{\rm{X}}$, the typical value of $\Gamma_{\rm{X}}=2.34$ for HBLs \citep{Donato2001hard} is adopted. The intrinsic 2--10\,keV X-ray luminosities ($L_{\rm{X,int}}$) are corrected from the observed ones ($L_{\rm{X,obs}}$) by applying a correction based on the X-ray Doppler factor ($\delta_{\rm{X}}$, see \S\,\ref{Beaming}).

\subsection{The BH Mass}
In this work, the dynamic BH masses ($M_{\rm{BH,dyn}}$) of HBLs are adopted for the FP, which are taken from the existing literature. As is well known, HBLs lack emission lines, hindering the virial estimation of their BH mass \citep{Shen2011}. Consequently, $M_{\rm BH,dyn}$ of HBLs are typically inferred from properties of their host galaxies, e.g., the $M_{\rm{BH}}-\sigma$ and $M_{\rm{BH}}-L_{\rm{bulge}}$ relation, where $\sigma$ and $L_{\rm{bulge}}$ denote the stellar velocity dispersion and bulge luminosity of the host galaxy, respectively. Here, we prefer BH masses estimated from the $M_{\rm{BH}}-\sigma$ relation. The $M_{\rm{BH,dyn}}$ of our HBLs are primarily taken from \cite{Paliya2021central}, with only a few sources from \cite{Woo2002active}, \cite{Wu2009debeamed} and \cite{Falomo2003host}. For 1421+582, no $M_{\rm{BH,dyn}}$ was reported in previous work, however, we extracted its mean absolute 
$R$-band magnitude $M_{\rm{R}}=-24.23$ in \cite{O2005host} and calculated its BH mass $\log M_{\rm{BH,dyn}}=9.20\,M_{\odot}$ using the $M_{\rm{BH}}-L_{\rm{bulge}}$ relation from \cite{Graham2007black}. For 1722+119, the $V$-band magnitude $M_{V}=-23.6$ was obtained from \cite{Falomo1993AJ}, and its BH mass ($\log M_{\rm{BH,dyn}}=9.87\,M_{\odot}$) was derived using the $M_{\rm{BH}}-L_{\rm{bulge}}$ relation in \cite{Graham2007black}. In total, the main parameters of our HBLs are listed in Table \ref{table2}, and the distribution of properties is shown in Fig.\,\ref{HIST}.

\begin{figure}
    \vspace{0.1cm}
    \centering{\includegraphics[scale=0.29]{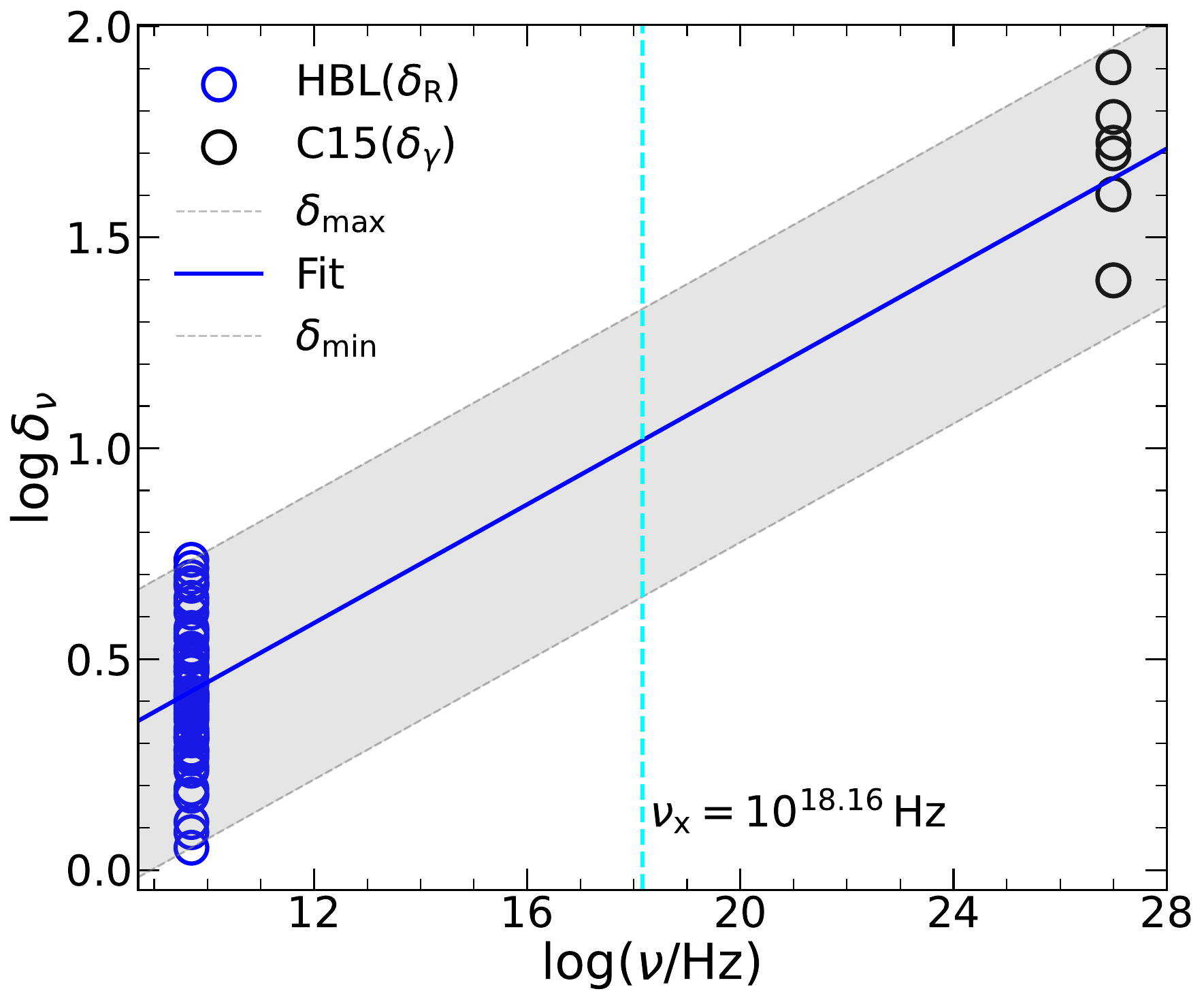}}
    \caption{The Doppler factor ($\delta_{\nu}$) as the function of frequency ($\nu$), which is based on the hypothesis of decelerating jet for HBLs \citep{Georganopoulos03,Ghisellini2005}. The blue circles are the 5\,GHz radio Doppler factors of our HBLs; the black circles are the typical TeV band ($10^{27}$\,Hz) Doppler factors for HBLs, which are taken from \citet{Cerruti2015}. The blue solid line is the best fit (Eq\,\ref{Eq3}); the top and bottom gray dashed line represent the intercept that derived from maximum and minimum $\delta_{\rm{R}}$, respectively.}
    \vspace{0.1cm}
    \label{D-v}
\end{figure}

\section{METHODS AND RESULTS}\label{METHODS AND RESULTS}

It is widely accepted that blazars have the Doppler beaming effect, because they exhibit a jet towards us \citep{Urry1995}. Therefore, the Doppler beaming effect should be considered for our HBLs even if they are weak jet sources, which allows us to explore whether the Doppler beaming effect has a significant impact on radio--X-ray correlation and FP for HBLs.

\subsection{The Doppler Beaming Effect}\label{Beaming}

In the Doppler beaming effect model, the observational flux density ($F_{\rm \nu,obs}$) is strongly boosted from the intrinsic flux density ($F_{\rm \nu,int}$):
\begin{equation}
   F_{\nu,\rm{int}}=F_{\nu,\rm{obs}}\delta_{\nu}^{-q}
    \label{Eq2}
\end{equation}
where $q = 2 + \alpha_{\nu}$, which corresponds to the continuous jets, and $\alpha_{\nu}$ is the spectral index. We used the typical value of $\alpha_{\rm{R}}=0$ \citep{Donato2001hard,Long.Q-C2025FP} for radio band, while the X-ray spectral indices are calculated using the $\alpha_{\rm{X}}=\Gamma_{\rm{X}}-1$ \citep{Lusso2010x}.

The Doppler factor is typically determined by the velocity ($\beta \sim v/c$, $c$ is speed of light) and the viewing angle ($\theta$) between the jet axis and the line of sight to the observer: $\delta=[\Gamma(1-\beta cos\theta)]^{-1}$, where $\Gamma$ is the Lorentz factor ($\Gamma=1/\sqrt{1-\beta^2}$). Considering the decelerating jet for HBLs \citep{Georganopoulos03,Ghisellini2005}, the different band Doppler factors ($\delta_{\nu}$) should be different. In this work, we only have the available 5\,GHz radio Doppler factor ($\delta_{\rm{R}}$), while the 2--10\,keV X-ray Doppler factors ($\delta_{\rm{X}}$) are absent. Therefore, we will attempt to explore the relationship between $\delta_{\nu}$ and the observational frequency ($\nu$) to obtain $\delta_{\rm{X}}$.

The bright and rapidly variable TeV emissions were observed in TeV-HBLs, implying that the gamma-ray emission region must be highly relativistic, with a large $\gamma$-ray Doppler factor ($\delta_\gamma$) \citep{Ghisellini2005,Cerruti2015}. However, VLBI radio studies indicate that pc-scale jets of TeV-HBLs move slowly \citep{Ghisellini2005,Piner2008parsec}, which suggests that the jet of HBLs may suffer severe deceleration from the $\gamma$-ray emitting region ($\sim0.1$\,pc from jet apex) to the VLBI scale \citep[$\sim1$\,pc, see][]{Ghisellini2005}. Therefore, we assume $\delta_{\nu}$ decreases with decrease of $\nu$. Considering that the jet starts to decelerate from the gamma-ray emission region, we taken the typical TeV band ($10^{27}$\,Hz) Doppler factor ($\delta_\gamma$) from \cite{Cerruti2015}, then combine them with $\delta_{\rm{R}}$ for our HBLs. By applying the orthogonal distance regression, we obtain the correlation between $\delta_{\nu}$ and $\nu$ (see Fig.\,\ref{D-v}):
\begin{equation}
    \log \delta_{\nu}=(0.070\pm0.003)\log \nu-(0.258\pm0.042)
    \label{Eq3}
\end{equation}
the blue solid line in Fig.\,\ref{D-v} represents Equation\,\ref{Eq3}, we can regard it as an average property of the Doppler factor. It is clear that each source corresponds to each intercept. Hence, we fix the slope $k=0.07$, then each point ($\log \delta_{\rm{R}}$, $\log 5\,\rm{GHz}$) can derive each intercept. Finally, we can obtain
\begin{equation}
    \delta_{\nu}=10^{(\log \delta_{\rm{R}}-0.682)}\nu^{0.07}
    \label{Eq4}
\end{equation}
Substituting the 2--10\,keV X-ray frequency ($\nu_{\rm{X}}=10^{18.16}\rm Hz$) and $\delta_{\rm R}$ into Equation\,\ref{Eq4}, we can get $\delta_{\rm{X}}$ for our HBLs (see the vertically cyan line in Fig.\,\ref{D-v}).

It is note that the observational X-ray emissions may be sauperposition of jet component and disk component \citep{Ghisellini2016,Bariuan2022fundamental,Wang2024FP,Long.Q-C2025FP}. \cite{Ghisellini2016} have suggested that there are $\sim20\%$ disk component in the SED of blazars (see their fig.\,3). Therefore, we assume that there are $\sim80\%$ jet components in the observational X-ray emissions (i.e., $F_{\rm{X,Jet}}=0.8F_{\rm{X,obs}}$), while the $\sim20\%$ disk components do not need to perform the Doppler correction, and we have
\begin{equation}
    F_{\rm{X,int}}=(0.8F_{\rm{X,obs}})\delta_{\rm{X}}^{-q}+0.2F_{\rm{X,obs}}
    \label{Eq5}
\end{equation}

\subsection{RESULTS}\label{result}
We performed the fit between the Eddington-luminosity-scaled radio and X-ray luminosity for our HBL sample in the form:
\begin{equation}
    \log (\frac{L_{\rm{R}}}{L_{\rm{Edd}}})=\xi_{\rm{RX}}\log (\frac{L_{\rm{X}}}{L_{\rm{Edd}}})+\rm{constant}
    \label{Eq6}
\end{equation}

To derive the FP for our HBL sample, we define the FP equation that is similar to \cite{Merloni2003} as the form:
\begin{equation}
    \log L_{\rm{R}}=\xi_{\rm{RX}}\log L_{\rm{X}}+\xi_{\rm{RM}}\log M_{\rm{BH}}+c
    \label{Eq7}
\end{equation}
To find the multivariate relation coefficients of FP, we adopted the least $\chi^2$ approach in \cite{Merloni2003} and minimized the following statistic:
\begin{equation}
    \chi^{2}=\sum\limits_{i}\frac{(y_i-c_0-\xi_{\rm RX}X_i-\xi_{\rm RM}M_i)^2}{\sigma^2_{\rm{R}}+\xi_{\rm RX}^2\sigma^2_{\rm X}+\xi_{\rm RM}^2\sigma^2_{\rm M}}
    \label{Chi-square}
\end{equation}
where the $y_i=\log L_{\rm R}$, $X_i=\log L_{\rm X}$, $M_i=\log M_{\rm BH}$ and $c_0$ is constant,  $\sigma_{\rm R}$, $\sigma_{\rm X}$, and $\sigma_{\rm M}$ are the associated error estimates of $\log L_{\rm R}$, $\log L_{\rm X}$, and $\log M_{\rm BH}$, respectively. As we all know, blazars typically exhibit the dramatic variability. However, the extreme variability is notoriously difficult to avoid for blazars. In particular, HBLs exhibit extreme variability on timescales as short as a few minutes \citep{Piner2010ApJ}. Therefore, the uncertainties of the data mainly come from the non-simultaneity between radio and X-ray emissions. To mitigate its effects, it is essential to take uncertainties into account when performing the linear regression. Considering the systematic and the observed uncertainties, we adopt the typical isotropic uncertainties with $\sigma_{\rm{R}}=\sigma_{\rm{X}}=0.3$\,dex \citep[see][]{Merloni2003,Xie2017fundamental,Long.Q-C2025FP}. In this work, the BH mass of HBLs are estimated using the $M_{\rm{BH}}-\sigma$ and $M_{\rm{BH}}-L_{\rm{bulge}}$ relation that derived from nearby galaxies. Therefore, the larger uncertainty of BH mass ($\sigma_{\rm{M}}=0.4\,\rm dex$) is adopted \citep{dong2014new}.

\begin{figure}
    \vspace{0.1cm}
    \centering{\includegraphics[scale=0.27]{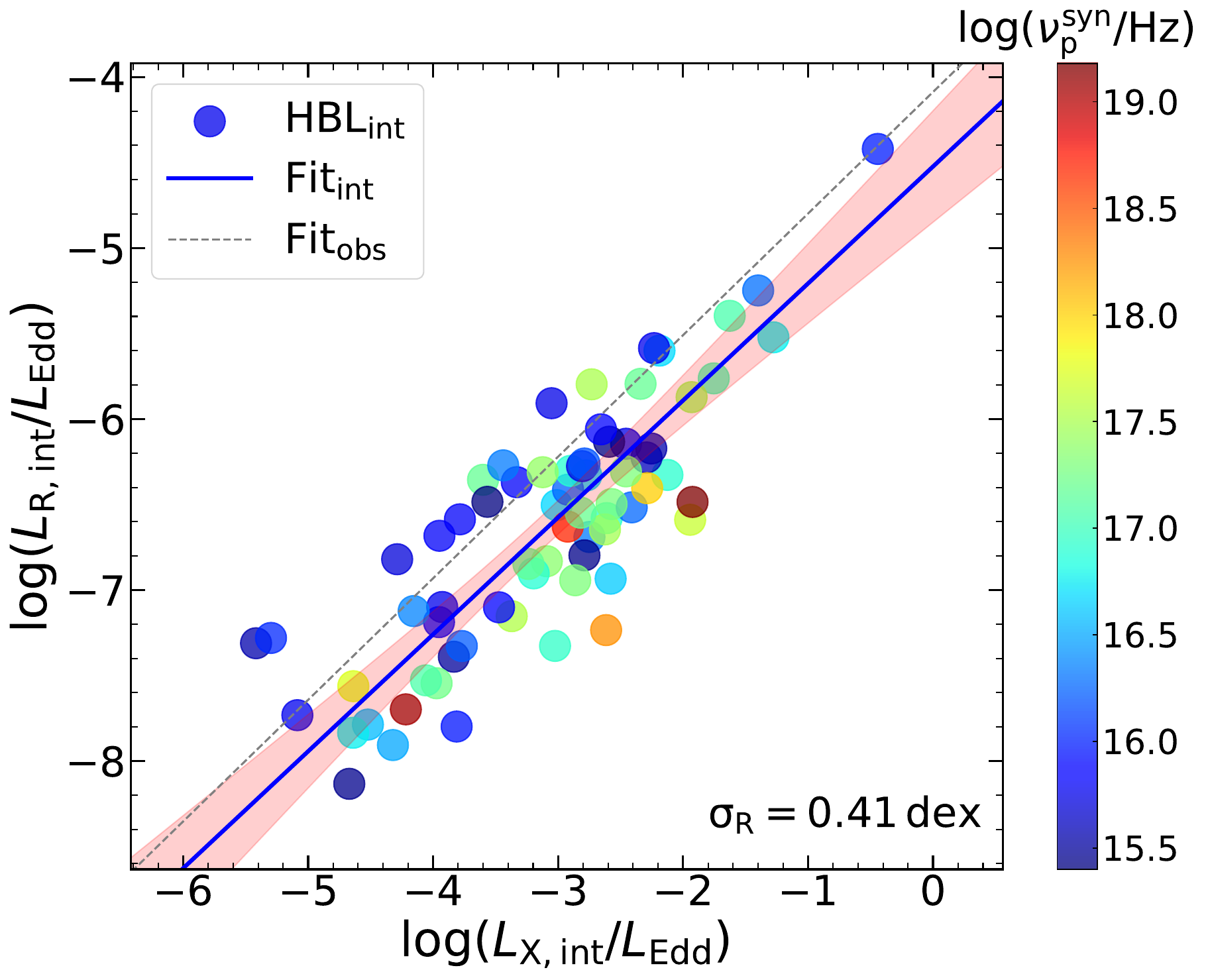}}
    \caption{The observational and intrinsic radio--X-ray correlation for HBLs. The 'int' and 'obs' indicate the observational data and the intrinsic data, respectively. The solid circles represent intrinsic data points of HBLs. The gray dashed line and the blue solid line are the best-fit of observational and intrinsic radio--X-ray correlation for HBLs. The the red shaded region is the $95\%(2\sigma)$ confidence intervals on the intrinsic fit. The scale of color denotes the size of $\rm log(\nu_p^{syn}/Hz)$ for our HBLs}.
    \vspace{0.1cm}
    \label{Lx/Ledd--Lr/Ledd}
\end{figure}

By considering the Doppler beaming effect, we obtain the intrinsic radio and X-ray luminosity for our HBLs. Through Eq\,\ref{Chi-square}, we obtain the observational and intrinsic radio--X-ray correlation and FP for HBLs. Fig.\,\ref{Lx/Ledd--Lr/Ledd} shows the Eddington-scaled radio--X-ray correlations for HBLs. The mathematical expressions of observational and intrinsic radio--X-ray correlation of HBLs are given as follows:

\begin{equation}
    \log (\frac{L_{\rm{R,obs}}}{L_{\rm{Edd}}})=(0.71\pm0.05)\log (\frac{L_{\rm{X,obs}}}{L_{\rm{Edd}}})-(4.09\pm0.11)
    \label{Eq9}
\end{equation}
with a scatter $\sigma_{\rm{R}}=0.38$\,dex, a Spearman correlation coefficient $R=0.84$ and $P$ value $P=4.86\times 10^{-19}$.
\begin{equation}
    \log (\frac{L_{\rm{R,int}}}{L_{\rm{Edd}}})=(0.68\pm0.05)\log (\frac{L_{\rm{X,int}}}{L_{\rm{Edd}}})-(4.52\pm0.15)
    \label{Eq10}
\end{equation}
with a scatter $\sigma_{\rm{R}}=0.41$\,dex, a Spearman correlation coefficient $R=0.79$ and $P$ value $P=5.17\times 10^{-16}$.

\begin{figure}
    \vspace{0.1cm}
    \centering{\includegraphics[scale=0.25]{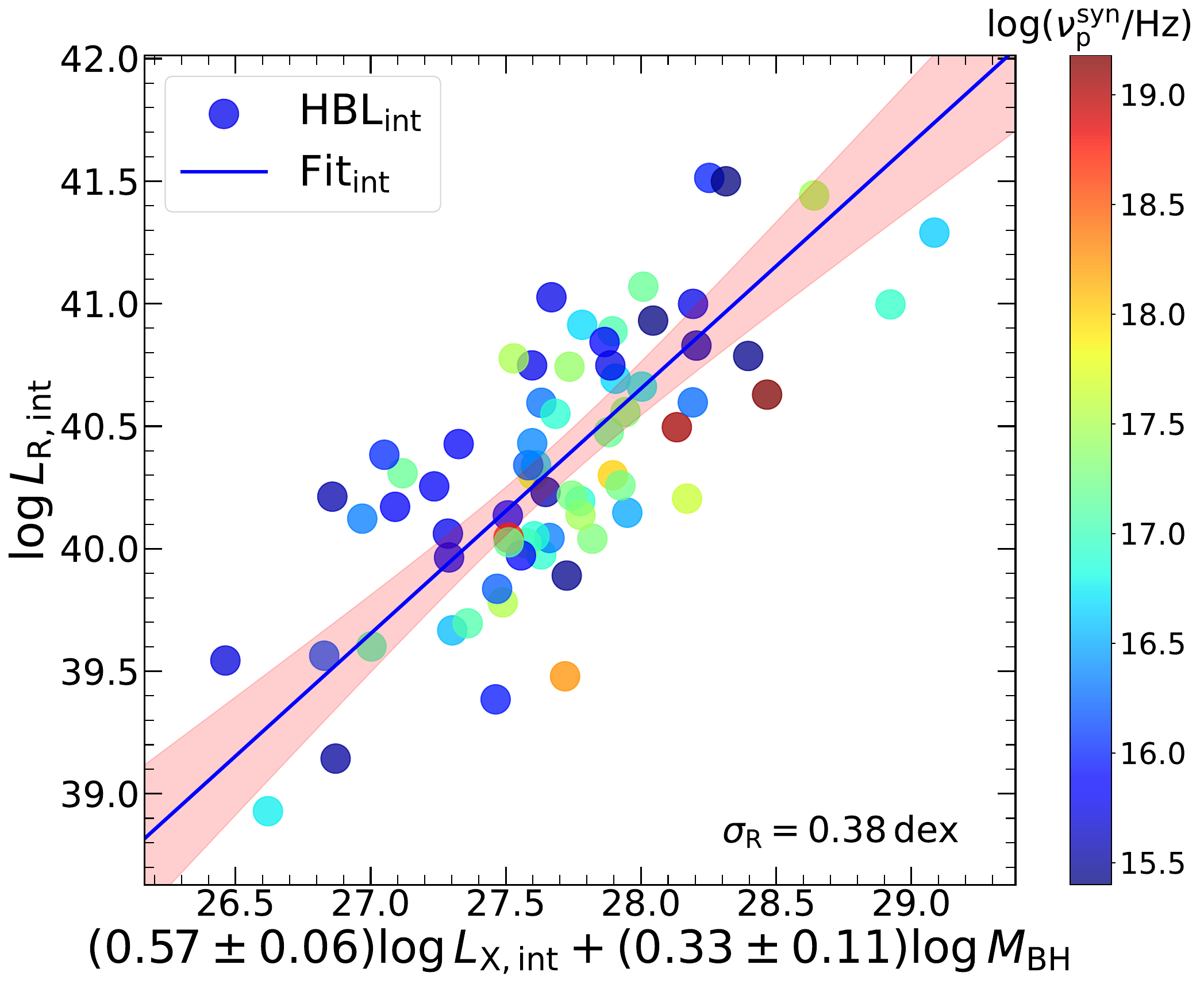}}
    \caption{The intrinsic FP for our HBLs. The notations are the same as that of Fig.\,\ref{Lx/Ledd--Lr/Ledd}.}
    \vspace{0.1cm}
    \label{FP}
\end{figure}

Fig.\,\ref{FP} shows the $\rm{FP_{int}}$ for HBLs. The mathematical expressions of observational FP ($\rm{FP_{obs}}$) and intrinsic FP ($\rm{FP_{int}}$) of HBLs are
\begin{equation}
    \begin{split}
        \log L_{\rm{R,obs}}=&(0.60\pm0.06)\log L_{\rm{X,obs}}+(0.25\pm0.10)\log M_{\rm{BH}}\\
        &+(12.07\pm1.96)
    \end{split}
    \label{Eq11}
\end{equation}
with a scatter $\sigma_{\rm{R}}=0.35$\,dex.
\begin{equation}
    \begin{split}
        \log L_{\rm{R,int}}=&(0.57\pm0.06)\log L_{\rm{X,int}}+(0.33\pm0.11)\log M_{\rm{BH}}\\
        &+(12.65\pm2.00)
    \end{split}
    \label{Eq12}
\end{equation}
with a scatter $\sigma_{\rm{R}}=0.38$\,dex.

A significant fraction of HBLs have featureless optical spectra ($\rm{EW}<5\text{\AA}$), which makes it challenging to measure their redshifts. Indeed, around 40\% of BL Lacs in the Fourth LAT AGN Catalog Data Release 2 lack measured redshifts \citep{Ajello2020ApJ,Sheng2024ApJ}. Although redshifts are available for our HBLs, some have considerable uncertainty, which raises the concern that they may introduce spurious results. However, the radio--X-ray correlation and FP do not depend on the redshifts \citep[e.g.,][]{Merloni2003,Li2008black,Bariuan2022fundamental,Wang2024FP,Long.Q-C2025FP}. Therefore, measurements of redshifts do not have a significant effect on our results. Following \cite{Long.Q-C2025FP}, we perform the Spearman partial correlation analysis to test the radio--X-ray correlation. For the given three variables ($X$, $Y$, $Z$), the partial correlation coefficient between $X$ and $Y$, while keeping $Z$ as the third variable, can be expressed as
\begin{equation}
    R_{XY,Z}=\frac{R_{XY}-R_{XZ}R_{YZ}}{[(1-R_{XZ}^2)(1-R_{YZ}^2)]^{1/2}}
\end{equation}
where the $R_{XY}$ denotes the Spearman rank correlation coefficient between $X$ and $Y$, and so on. Table\,\ref{table1} shows the results of our test, namaly, whether the correlation between $X$ and $Y$ is intrinsic or only introduced by a third variable $Z$. The null hypothesis will be rejected when its probability is less than the significance level (i.e., 0.05).

\begin{figure*}
    \centering
    \includegraphics[width=\textwidth]{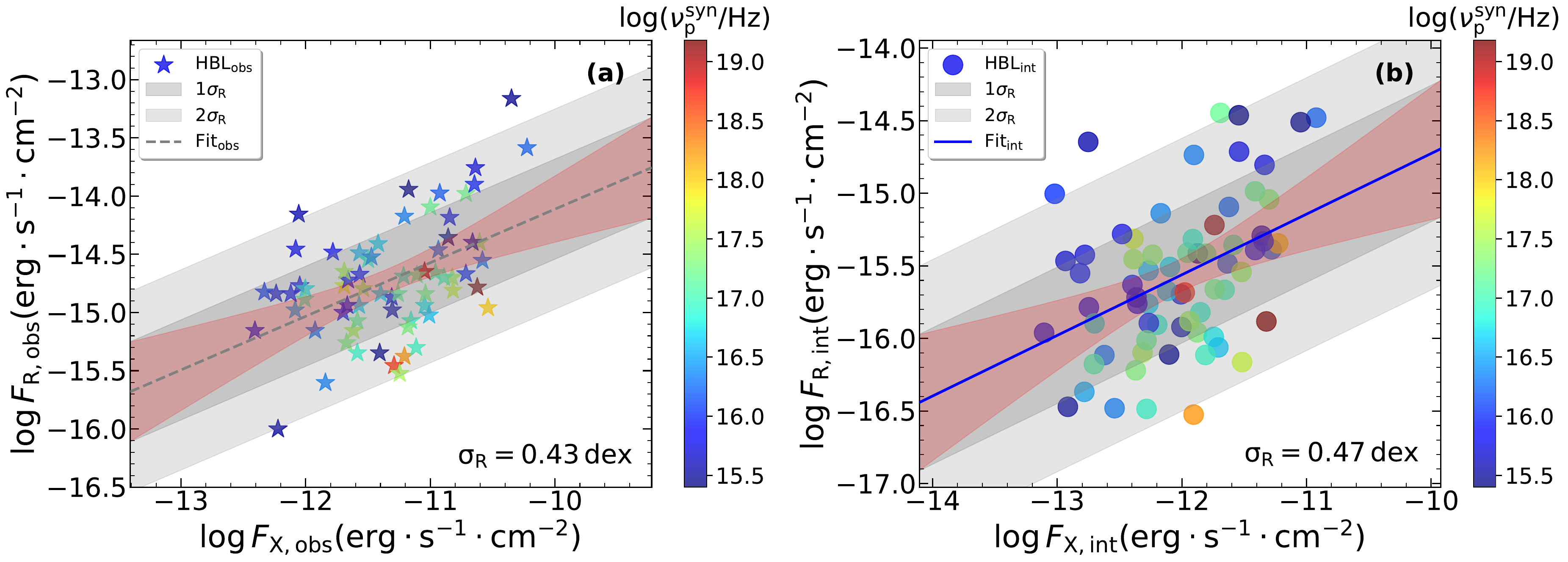}
    \caption{The observational and intrinsic correlations between radio and X-ray flux for our HBLs. The left panel is $\log F_{\rm R,obs}-\log F_{\rm X,obs}$ relation (the stars are the observational data points),  The right panel is $\log F_{\rm R,int}-\log F_{\rm X,int}$ relation. The notations are the same as that of Fig.\,\ref{Lx/Ledd--Lr/Ledd}. Shaded areas correspond to $1\sigma_{\rm R}$ and $2\sigma_{\rm R}$ (vertical) dispersion.}
    \label{Fr-Fx}
\end{figure*}

However, the significance level of radio--X-ray correlations become weaker when the distance is taken into account. As in \cite{Wang2006black}, in order to avoid the effect of distance, we test the existence of the intrinsic correlation between radio and X-ray emissions by comparing the radio and X-ray flux. However, it is clear that the ranges of variables are stretched when the distances are included, namely, the range of flux ($F_\nu$) is significantly shorter than that of luminosity ($L_\nu$). \cite{Long.Q-C2025FP} suggested that, in linear regression analysis, the range of variable should ideally be greater than 3 (i.e., $\Delta \log F_\nu=\log F_{\rm\nu,max}-\log F_{\rm\nu,min}\gtrsim3$); otherwise, it may lead to misleading results. When the range of a variable is relatively small, the ordinary least squares (OLS) appears to be a better regression method \citep[e.g.,][]{Wang2024FP}. By employing the OLS, we obtain the observational and intrinsic correlations between radio and X-ray flux as the follows:
\begin{equation}
    \log F_{\rm R,obs}=(0.46\pm0.10)\log F_{\rm X,obs}-(9.51\pm1.13)
\end{equation}
with a scatter $\sigma_{\rm R}=0.43 \rm \,dex$, a Spearman correlation coefficient $R=0.44$ and $P=1.36\times10^{-4}$.
\begin{equation}
    \log F_{\rm R,int}=(0.42\pm0.11)\log F_{\rm X,int}-(10.55\pm1.32)
\end{equation}
with a scatter $\sigma_{\rm R}=0.47 \rm \,dex$, a Spearman correlation coefficient $R=0.42$ and $P=3.09\times10^{-4}$. Fig.\,\ref{Fr-Fx} shows our results. It is clear that the shallow correlations between radio and X-ray emissions still exist. Moreover, $\sigma_{\rm R}$ of $F_{\rm R}-F_{\rm X}$ relations is only slightly larger than that of $L_{\rm R}-L_{\rm X}$ relations (see Fig.\,\ref{Lx/Ledd--Lr/Ledd} and \ref{Fr-Fx}), but the significant level ($R$, $P$) of $F_{\rm R}-F_{\rm X}$ relations is significantly less than that of $L_{\rm R}-L_{\rm X}$ relations (see Table\,\ref{table1}), which is due to the distance effect stretching the range of variable \citep[e.g.,][]{Wang2006black,Long.Q-C2025FP}.

\begin{table}
    \caption{The Spearman Partial Correlation Analysis}
    \begin{tabular}{cccccc}
        \hline
        \hline
        $X$ & $Y$ & $Z$ & $R_{XY,Z}$ & $P_{\rm null}$\\
        \hline
        $\log \frac{L_{\rm X,obs}}{L_{\rm Edd}}$ & $\log \frac{L_{\rm R,obs}}{L_{\rm Edd}}$ & None & 0.84 & $4.86\times10^{-19}$\\
        $\log \frac{L_{\rm X,obs}}{L_{\rm Edd}}$ & $\log \frac{L_{\rm R,obs}}{L_{\rm Edd}}$ & $\log D$ & 0.81 & $4.41\times10^{-17}$\\
        $\log \frac{L_{\rm X,int}}{L_{\rm Edd}}$ & $\log \frac{L_{\rm R,int}}{L_{\rm Edd}}$ & None & 0.79 & $5.17\times10^{-16}$\\
        $\log \frac{L_{\rm X,int}}{L_{\rm Edd}}$ & $\log \frac{L_{\rm R,int}}{L_{\rm Edd}}$ & $\log D$ & 0.77 & $1.01\times10^{-14}$\\
        $\log F_{\rm X,obs}$ & $\log F_{\rm R,obs}$ & None & 0.44 & $1.36\times10^{-4}$\\
        $\log F_{\rm X,int}$ & $\log F_{\rm R,int}$ & None & 0.42 & $3.09\times10^{-4}$\\
        \hline
    \end{tabular}
    \label{table1}\\
    \raggedright{$\log D$ is the distance in units of Mpc}\\
\end{table}

The $\rm \nu_p^{syn}$ can be considered as the indicator of the energy distribution. It is worthy to investigate that whether $\rm \nu_p^{syn}$ has a significant effect on radio--X-ray correlation and FP. Therefore, we confer our HBLs in Fig.\,\ref{Lx/Ledd--Lr/Ledd}$-$\ref{Fr-Fx} with $\rm log(\nu_p^{syn}/Hz)$ values in the scale of color. It is clear that the distribution of $\rm \nu_p^{syn}$ shows a chaotic behavior, which appears to indicate $\rm \nu_p^{syn}$ has no a significant effect on the radio--X-ray correlation and FP for HBLs. However, a larger HBL sample and better simultaneous observational data are required to further test it in the future.

\section{Discussion}\label{Discussion}
\subsection{A New Fundamental Plane Derived From Synchrotron Cooling}\label{New-FP}

As shown in section\,\ref{result}, the observational and intrinsic radio--X-ray correlation of our HBLs are $L_{\rm R,obs}\propto L_{\rm X,obs}^{0.71}$ and $L_{\rm R,int}\propto L_{\rm X,int}^{0.68}$, respectively. The fitting coefficient of observational and intrinsic FP are $\xi_{\rm RX}=0.60$ and $\xi_{\rm RX}=0.57$, respectively. Our results are roughly consistent with $L_{\rm R}\propto L_{\rm X}^{0.64}$ reported by \cite{Donato2005six}. This agreement indicates that our results are not artifacts. These results suggest two possible origins of X-ray emissions, namely, ADAF-dominated mode and synchrotron cooling. For HBLs, the X-ray band is located at the hump of synchrotron emissions \citep{Fossati1998unifying,Donato2001hard}, which implies that the X-ray emissions of HBLs are dominated by synchrotron process of jets (the other components may be swamped by jet components). In addtion, the recent X-ray polarization observations of HBLs imply that X-ray emissions of HBLs may from synchrotron cooling \citep{Di-Gesu2023,Errando2024,Pacciani2025}. Therefore, we think that the synchrotron cooling origin of X-ray emissions is the more logical explanation for our results. In the follows, we will discuss our analysis based on the theoretical model of synchrotron cooling in \cite{Heinz2004}. 

\cite{Heinz2004} have suggested that the X-ray emissions from synchrotron cooling can reproduce the similar FP ($\xi_{\rm RX}=0.6$) found by \cite{Merloni2003}. Considering the effects of radiative cooling on the X-ray synchrotron emission, \cite{Heinz2004} derived the desired relation between the radio flux ($F_{\rm{R}}$), the BH mass ($M_{\rm{BH}}$) and the synchrotron X-ray flux ($F_{\rm{X}}$):
\begin{equation}
   F_{\rm{R}}\propto F_{\rm{X}}^{\xi_{\rm{RX}}} M_{\rm{BH}}^{\xi_{\rm{RM}}}
\end{equation}
\begin{equation}
    \xi_{\rm{RX}}=\frac{2p+13+(p+6)\alpha_{\rm{R}}}{(p+4)(2p+1-3\alpha_{\rm{X}})}
    \label{Eq17}
\end{equation}
\begin{equation}
    \xi_{\rm{RM}}=\frac{(2p+13-(2+p)\alpha_{\rm{R}})(p-1-\alpha_{\rm{X}})-2\alpha_{\rm{R}}}{(p+4)(2p+1-3\alpha_{\rm{X}})}
    \label{Eq18}
\end{equation}
where the $p$ is the power-law index of the accelerated relativistic electrons (${\rm{d}}N_{e}/{\rm{d}}\gamma \sim \gamma^{-p}$, $\gamma_{\rm{min}}<\gamma<\gamma_{\rm{max}}$, $\gamma$ is the particle Lorentz factor), $\alpha_{\rm{R}}$ is the radio spectral index and $\alpha_{\rm{X}}$ is the X-ray spectral index. It is clear that the correlation coefficients ($\xi_{\rm{RX}}$, $\xi_{\rm{RM}}$) of FP are determined by $p$, $\alpha_{\rm{R}}$, and $\alpha_{\rm{X}}$. 

In the canonical synchrotron radiation, the radiative cooling is negligible for the region of the jet where the X-rays are produced, leaving $p\sim2$ over the entire spectrum \citep{Merloni2003,Heinz2003,Heinz2004}. In this case, the X-ray spectrum should continue from the X-ray band to lower frequencies at the standard synchrotron spectral, i.e, $\alpha_{\rm{X}}=(p-1)/2$ \citep[see][]{Merloni2003,Heinz2004}. However, the synchrotron cooling can produce a universal power-law with a slope of order $p>3$ \citep[see][]{Merloni2003,Heinz2004,Kino2002ApJ,Kino&Takahara2004,Uhm&Zhang2014}, which corresponds to $\alpha_{\rm{X}}>1$. Therefore, we can summarize as follows:
\begin{equation}
        {\rm{d}}N_{e}/{\rm{d}}\gamma \sim \gamma^{-p}
        \begin{cases}\rm{\text{C-syn}}&p\sim2,\alpha_{\rm{X}}\sim0.5\\\rm{\text{Syn-c}}&p>3,\alpha_{\rm{X}}>1\end{cases}
\end{equation}
\cite{Heinz2004} found that the scenario of synchrotron cooling ($p\sim3.4$, $\alpha_{\rm{X}}=1.2$, $\alpha_{\rm{R}}=0$) can reproduce the FP found by \cite{Merloni2003}.

\begin{figure}
    \vspace{0.1cm}
    \centering{\includegraphics[scale=0.292]{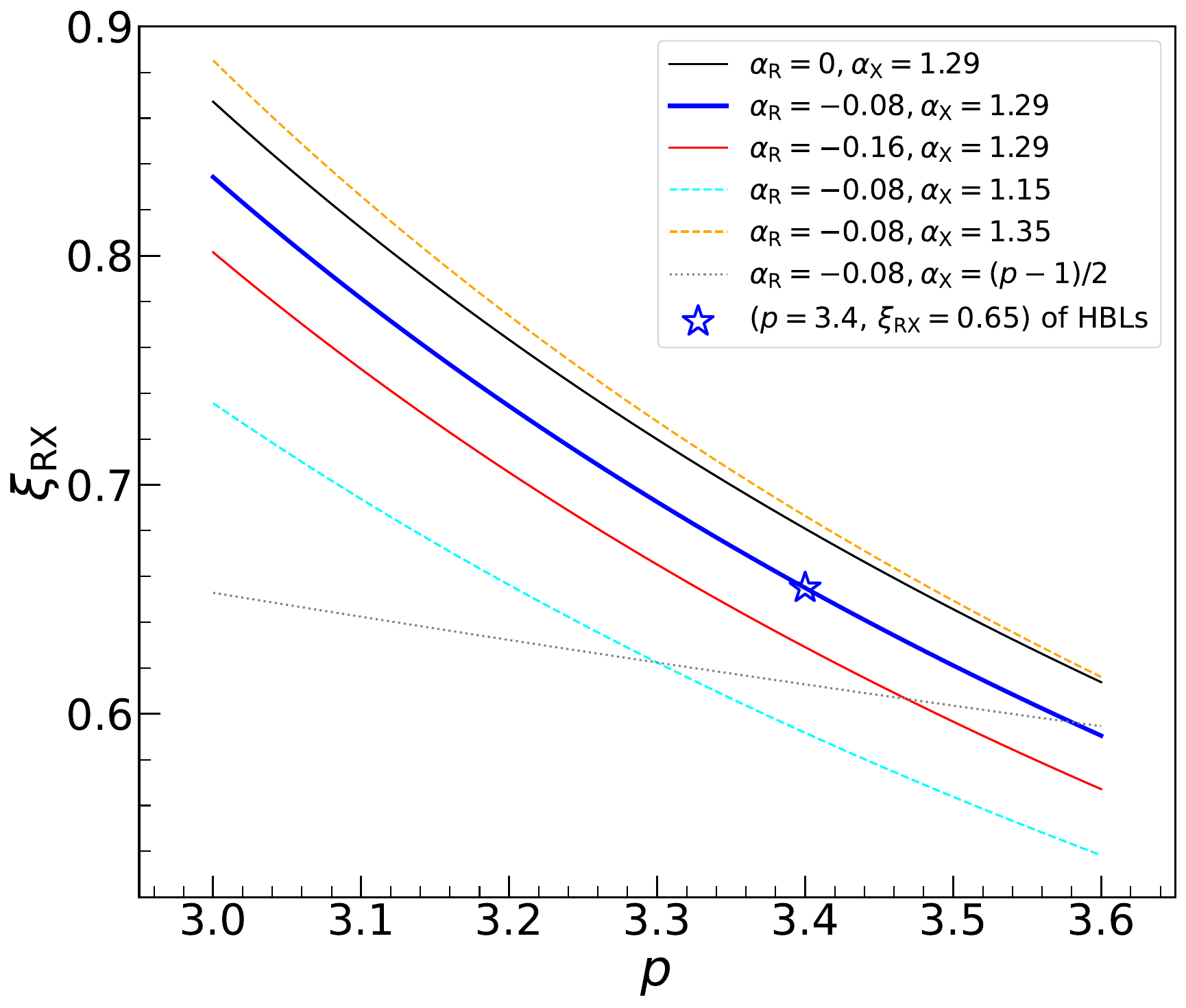}}
    \caption{The black solid line, blue solid line and red solid line are the $\xi_{\rm{RX}}-p$ functions under the fixed parameters of ($\alpha_{\rm{R}}=0$, $\alpha_{\rm{X}}=1.29$), ($\alpha_{\rm{R}}=-0.08$, $\alpha_{\rm{X}}=1.29$) and ($\alpha_{\rm{R}}=-0.16$, $\alpha_{\rm{X}}=1.29$), respectively. The orange dashed line and the cyan dashed line are the $\xi_{\rm{RX}}-p$ functions under the fixed parameters of ($\alpha_{\rm{R}}=-0.08$, $\alpha_{\rm{X}}=1.15$) and ($\alpha_{\rm{R}}=-0.08$, $\alpha_{\rm{X}}=1.35$), respectively. The gray dotted line is the $\xi_{\rm{RX}}-p$ function under the assumption \citep[$\alpha_{\rm{R}}=-0.08$, $\alpha_{\rm{X}}=(p-1)/2$, see][]{Heinz2004}. The blue star is the theoretical prediction ($\xi_{\rm{RX}}=0.65$) of our HBLs by taking typical values ($p=3.4$, $\alpha_{\rm{R}}=-0.08$, $\alpha_{\rm{X}}=1.29$) into account.}
    \vspace{0.1cm}
    \label{p-value}
\end{figure}

Considering the X-ray emissions of HBLs originate from synchrotron cooling \citep{Di-Gesu2023}, the $p$ value of our HBLs should be lager than 3 (i.e., $p>3$). By self-consistently taking into account the effects of radiative cooling, \cite{Kino2002ApJ} numerically calculated the steady state energy spectra of electrons for HBLs, they found that radiative cooling decreases the number density of electrons and leads to a break in the relativistic electron energy spectrum (see their fig.\,2), yielding $p>3$ \citep[also see][]{Kino&Takahara2004}. In the following, we will adopt the average value of $\langle\alpha_{\rm{X}}\rangle=1.29$ for our HBLs. For the radio spectral index of HBLs, we adopt the average value of $\langle\alpha_{\rm{R}}\rangle=-0.08$ from \cite{Fuhrmann2016}. We substitute $p=3.4$, $\alpha_{\rm{X}}=1.29$ and $\alpha_{\rm{R}}=-0.08$ into Eq\,\ref{Eq17} and Eq\,\ref{Eq18}, then obtain $\xi_{\rm{RX}}\sim0.65$ and $\xi_{\rm{RM}}\sim0.78$, which roughly agree with the intrinsic radio--X-ray correlation ($\xi_{\rm{RX}}\sim0.68$) and intrinsic FP ($\xi_{\rm{RX}}\sim0.57$), suggesting that the X-ray emissions of HBLs are dominated by the synchrotron cooling.

In order to explore the dependence of $\xi_{\rm{RX}}$ on the parameter $p$, $\alpha_{\rm{X}}$ and $\alpha_{\rm{R}}$, we fix $\alpha_{\rm{X}}$ and $\alpha_{\rm{R}}$ as a constant, then fit the function of $\xi_{\rm{RX}}$ against the $p$ value. Subsequently, we investigate the effect of $\alpha_{\rm{X}}$ and $\alpha_{\rm{R}}$ on the $\xi_{\rm{RX}}-p$ function. Considering the $\alpha_{\rm{X}}$ and $\alpha_{\rm{R}}$ as the average value of our HBLs, we predict that $\alpha_{\rm{X}}$ and $\alpha_{\rm{R}}$ would not change too much. Therefore, we assume $-0.16<\alpha_{\rm{R}}<0$ and $1.15<\alpha_{\rm{R}}<1.35$, the typical value of $\alpha_{\rm{R}}=-0.15$ reported by \cite{Plotkin12} and $\langle\alpha_{\rm{X}}\rangle=1.34$ found by \cite{Donato2001hard} are also included in the range of our assumption.  Fig.\,\ref{p-value} shows our results, it is clear that $\xi_{\rm{RX}}$ is positively correlated with $\alpha_{\rm{R}}$ and $\alpha_{\rm{X}}$ (i.e., $\xi_{\rm{RX}}\propto \alpha_{\rm{R}}$ and $\xi_{\rm{RX}}\propto \alpha_{\rm{X}}$), the $\xi_{\rm{RX}}$ is negatively correlated with $p$ value (i.e., $\xi_{\rm{RX}}\propto 1/p$). Therefore, it is clear that $p>3$ cause the radio--X-ray correlation and FP to be shallower, which implies the X-ray emissions are produced by rapidly cooling, high-energy electrons accelerated at a shock \citep[see][]{Di-Gesu2023}. Therefore, the care must be taken when placing the beaming sources on diagnostic relations in \cite{Merloni2003} and \cite{yuan2005radio}. In the follow, we give a diagnostic scheme to avoid the misleading inference:
\begin{equation}
    L_{\rm R}\propto L_{\rm X}^{0.6\sim0.7}
    \begin{cases}\rm{ADAF}&p\sim2,\alpha_{\rm{X}}\sim0.5\\\rm{\text{Syn-c}}&p>3,\alpha_{\rm{X}}>1\end{cases}
\end{equation}

\subsection{The Impact of Doppler Beaming Effect on the Radio--X-ray Correlation and FP for Weak Jet Sources}

It is generally believed that blazars have the Doppler beaming effect due to they exhibit a relativistic jet towards us \citep{Urry1995,Fossati1998unifying,Yang2022beaming}. \cite{Long.Q-C2025FP} have suggested that the strong jet blazars ($\rm log(\nu_p^{syn}/Hz)<15.3$) have a nonnegligible Doppler beaming effect. However, the HBLs are weak jet sources \citep[e.g.,][]{Meyer2011blazar,Keenan2021relativistic} with a relative smaller Doppler factor (see Fig.\,\ref{HIST}c), and their Doppler factors are in the small range ($1.13\leq \delta \leq 5.44$), with an average value $\langle \delta_{\rm{R}}\rangle=2.81$, indicating that the most of HBLs have a near-equivalent and a weak Doppler boosting effect in luminosity ($\Delta L=L_{\rm{obs}}-L_{\rm{int}}$). Therefore, we can see that the gap between the blue solid line and the gray dashed line in Fig.\,\ref{Lx/Ledd--Lr/Ledd} is almost equal, in contrast to that of the strong jet blazars is getting bigger \citep[see fig.\,3 in][]{Long.Q-C2025FP}. From section \ref{result}, we can find that slope of the Doppler corrected radio--X-ray correlation and FP do not change significantly (also see Fig.\,\ref{Lx/Ledd--Lr/Ledd}), which imply that the Doppler beaming effect does not have a significant impact on the radio--X-ray correlations and FP for the weak jet sources.

There is a caution here, the Doppler factor is an unobserved quantity and a variable. This physical quantity is notoriously difficult to estimate as $\beta$ and $\theta$ are unobservable. Therefore, the $\delta$ of HBLs in this work may have a large error. Moreover, our $\delta_{\rm R}$ are taken from \cite{Wu2014some} and \cite{Ye2021unification}, and they are estimated by different methods, which may introduce an error into our results. However, we expect this to have no significant impact on our results. Because HBLs have a relatively small $\delta_{\rm R}$ with an average value $\langle \delta_{\rm{R}}\rangle=2.81$, and \cite{Wu2007} found that the $\delta-\nu_{\rm p}^{\rm syn}$ relation of HBLs shows an almost horizontal line, which may be due to the fact that HBLs are weak jet sources. Therefore, we can find that the weak relativistic effect do not have a significant impact on the radio--X-ray correlations and FP for HBLs (as the mention above).

In addition, the helical jet precession or the twisted inhomogeneous jet may cause the $\theta$ to be change, which causes the $\delta$ to be change \citep[$\delta$ is a variable, see][]{Raiteri2017blazar}. Therefore, the radio and X-ray data should be taken from the relative plateau of HBLs rather than its flare, and the mean value of radio and X-ray data should be used if their observations have multiple values. These measures can mitigate the effects of dramatic variability of HBLs on our results.

\subsection{Whether Our HBL Sample can be Representative of the Broader HBL Population?}
In the 3HSP sample \citep{Chang2019}, there are 2013 blazars with $\rm log(\nu_p^{syn}/Hz)>15$, which are usually HBLs \citep[see][]{Abdo2010ApJ,Giommi2012}. However, limited by our HBL selection criteria (i.e., HBLs have available 5\,GHz core radio flux, X-ray flux, BH mass, Doppler factor, and $\rm \nu_p^{syn}$), our final sample contains only 69 HBLs. In the following, we will compare some properties of our sample with a larger HBL sample from \cite{Wu2007}, \cite{Fan2016spectral}, and \cite{Chang2019}. This comparison will verify that our sample can be roughly regarded as representative of the broader HBL population. 

(1) As shown is Fig.\,\ref{HIST}(a), the range of redshift of our HBL sample is $0.030\le z \le 0.702$ with a mean $\langle z\rangle=0.206$. There are 1773 HBLs with available redshift in 3HSP sample \citep[see][]{Chang2019}, their redshift range is $0.000\le z \le 1.239$ with a mean $\langle z\rangle=0.363$, but HBLs with $z>0.702$ have a small fraction in 3HSP sample. Indeed, many previous studies are strikingly consistent in showing that HBLs are usually detected at lower redshifts \citep[$z<1$, e.g.,][]{Fossati1998unifying,Fan2016spectral,Yang2022spectral}.

(2) As shown in Fig.\,\ref{HIST}(b), the range of $\rm \nu_p^{syn}$ of our HBL sample is $\rm 15.40\le log(\nu_p^{syn}/Hz)\le 19.18$, the range of $\nu_{\rm p}^{\rm syn}$ of 3HSP sample is $\rm 15.00\le log(\nu_p^{syn}/Hz)\le 18.50$, the range of $\nu_{\rm p}^{\rm syn}$ of (207) HBL sample from \cite{Fan2016spectral} is $\rm15.30\le log(\nu_p^{\rm syn}/Hz)\le 19.29$. Note, \cite{Chang2019} defined BL Lacs with $\rm log(\nu_p^{\rm syn}/Hz)>15$ as HBLs. In this work, we adopt the classification criterion from \cite{Fan2016spectral}, which defines BL Lacs with $\rm log(\nu_p^{syn}/Hz) > 15.3$ as HBLs (see section\,\ref{Intro}). It is clear that our HBL sample nearly covers the range of $\rm \nu_{\rm p}^{syn}$ of the broader HBL population.

(3) As shown in Fig.\,\ref{HIST}(g), the range of $L_{\rm R,obs}$ of our HBL sample is $39.615\le \log L_{\rm R,obs}\le 42.402$ with a mean $\langle \log L_{\rm R,obs}\rangle=41.164$, which roughly agrees with those in \cite{Wu2007} (83 HBLs: $39.499\le \log L_{\rm R,obs}\le 42.809$ with a mean $\langle \log L_{\rm R,obs}\rangle=41.173$). Additionally, there are 123 HBLs with available redshift in HBL sample of \cite{Fan2016spectral}, they provided observational 1.4\,GHz total radio luminosity ($L_{\rm R,obs}^{\rm 1.4GHz}$: which is contaminated by extended radio emissions) for these HBLs, and the range of $L_{\rm R,obs}^{\rm 1.4GHz}$ of they HBL sample is $39.05\le \log L_{\rm R,obs}^{\rm 1.4GHz}\le 42.77$ (only one have $\log L_{\rm R,obs}^{\rm 1.4GHz}=43.48$) with a mean $\langle \log L_{\rm R,obs}^{\rm 1.4GHz}\rangle=41.071$. Despite the differences in the observational frequency (5\,GHz vs. 1.4\,GHz) and in the radio component (core vs. total), the $\log L_{\rm R,obs}$ range of our HBL sample is only slightly smaller than the $\log L_{\rm R,obs}^{\rm 1.4GHz}$ range of the HBL sample in \cite{Fan2016spectral}. It is clear that our HBL sample roughly covers the range of radio luminosity of the broader HBL population.

(4) As shown in Fig.\,\ref{HIST}(h), the range of $L_{\rm X,obs}$ of our HBL sample is $42.744\le \log L_{\rm X,obs}\le 46.343$ with a mean $\langle \log L_{\rm X,obs}\rangle=44.560$. The range of observational 1\,keV X-ray luminosity ($L_{\rm X,obs}^{\rm 1keV}$) of 123 HBLs from \cite{Fan2016spectral} is $42.12\le \log L_{\rm X,obs}^{\rm 1keV}\le 46.99$ with a mean $\langle \log L_{\rm X,obs}^{\rm 1keV}\rangle=44.648$. Despite the difference in the observational frequency (2--10\,keV vs. 1\,keV), the range of $L_{\rm X,obs}$ of our HBLs is slightly smaller than the range of $L_{\rm X,obs}^{\rm 1keV}$ and they have similar mean.

(5) In addition to classification that based on $\nu_{\rm p}^{\rm syn}$, BL Lacs are also divided into LBLs and HBLs according to the broad-band spectral index from radio band to X-ray band \citep[$\alpha_{\rm RX}$: $\alpha_{ij}=-\log(F_i/F_j)/\log(\nu_i/\nu_j)$, see][]{Ledden1985ApJ}. Generally, LBLs have $\alpha_{\rm RX}>0.75$, on the contrary, HBLs have $\alpha_{\rm RX}<0.75$ \citep[see][]{Padovani1995ApJ}. Fig.\,\ref{HIST}-(e) shows our result, it is clear that our HBLs have $\alpha_{\rm RX}\lesssim0.75$.

Taken together, these comparisons above verify that our sample can be roughly regarded as representative of the broader HBL population.

\section{SUMMARY}\label{SUMMARY}

The radio--X-ray correlation and FP are the practical diagnostics to constrain the accretion mode and the origin of X-ray emissions in BH system \citep[e.g.,][]{Merloni2003,yuan2005radio,Kording06RefiningA&A,Plotkin12,dong2014new,Wang2024FP,Long.Q-C2025FP}. However, a misleading radio--X-ray correlation ($L_{\rm{R}}\propto L_{\rm{X}}^{0.64}$) of HBLs was found by \cite{Donato2005six}, which can be explained by ADAF-dominated mode or synchrotron cooling. Both early observations and recent polarization measurements show that X-ray emissions of HBLs originate from the synchrotron process of jets \citep[e.g.,][]{Fossati1998unifying,Donato2001hard,Fan2016spectral,Di-Gesu2023,Errando2024,Pacciani2025}. Therefore, we think that the synchrotron cooling is more plausible explanation for this shallow relation. To further clarify the origin of the X-ray emissions of HBLs and explain the possible physics behind the shallow radio--X-ray correlation. We have compiled a sample of 69 HBLs with available core radio flux density, X-ray flux, dynamical BH mass, and Doppler factors in order to re-explore the radio--X-ray correlation and FP for HBLs. We find our fitting results are consistent with that of \cite{Donato2005six}. By employing the theoretical model of synchrotron cooling in \cite{Heinz2004}, we find that these shallow correlations are caused by synchrotron cooling. Our main results can be summarized as the follows:

(1) By considering the Doppler beaming effect, we obtain the intrinsic radio--X-ray correlations and FP for HBLs, they are $L_{\rm R,int}\propto L_{\rm X,int}^{0.68}$ and $\log L_{\rm R,int}=(0.57\pm0.06)\log L_{\rm X,int}+(0.33\pm0.11)\log M_{\rm BH}+(12.65\pm2.00)$, respectively. These results agree with the previous radio--X-ray correlation ($L_{\rm{R}}\propto L_{\rm{X}}^{0.64}$) found by \cite{Donato2005six}. By considering the synchrotron cooling, we substitute the typical values of synchrotron cooling ($p=3.4$, $\alpha_{\rm{X}}=1.29$, $\alpha_{\rm{R}}=-0.08$) of HBLs into the theoretical prediction of FP \citep[see Eq\,\ref{Eq17}, Eq\,\ref{Eq18} or][]{Heinz2004}, then obtain theoretical value of $\xi_{\rm{RX}}=0.65$, which roughly agrees with our intrinsic results, suggesting that the X-ray emissions of HBLs mainly originate from synchrotron cooling (section\,\ref{New-FP}). Our results provide the first observational evidence of $L_{\rm R}\propto L_{\rm X_{\text{Syn-c}}}^{0.6\sim0.7}$.

(2) Our results indicate that the Doppler beaming effect does not have a significant effect on the radio--X-ray correlation and FP for HBLs (weak jet sources).

\section*{Acknowledgements}

This work is supported by the NSFC (12363005), the National SKA Program of China (2022SKA0130104), the Foundation of Guizhou Provincial Education Department ((2020)0030), the Scientific Research Project of the Guizhou Provincial Education (KY[2022]132, KY[2022]137), Major Science and Technology Program of Xinjiang Uygur Autonomous Region (2022A03013-4) and Projects of the Grassroots Science Popularization Action Plan of Guizhou Provincial Association for Science and Technology.

\clearpage
\onecolumn

\begin{landscape}
\centering
\begin{longtable}
                {ccccccccccccccc}
\caption{The Properties of HBLs}\label{table2} \\
\toprule
IAU Name & RA & Dec & $z$ & $\log (\nu_{\rm p}^{\rm syn}/\rm{Hz})$ & $\delta_{\rm R}$ & 
$\Gamma_{\rm X}$ & Refs. & $\log L_{\rm X,obs}$ & $\log L_{\rm X,int}$ & $\log L_{\rm R,obs}$ & $\log L_{\rm R,int}$ & $\alpha_{\rm RX}$ & $\log M_{\rm BH,dyn}$ & $\lambda_{\rm int}$ \\ 
(1) & (2) & (3) & (4) & (5) & (6) & (7) & (8) & (9) & (10) & (11) & (12) & (13) & (14) & (15) \\
\midrule
\endfirsthead

\caption[]{The Properties of HBLs (continued)} \\
\toprule
IAU Name & RA & Dec & $z$ & $\log (\nu_{\rm p}^{\rm syn}/\rm{Hz})$ & $\delta_{\rm R}$ & 
$\Gamma_{\rm X}$ & Refs. & $\log L_{\rm X,obs}$ & $\log L_{\rm X,int}$ & $\log L_{\rm R,obs}$ & $\log L_{\rm R,int}$ & $\alpha_{\rm RX}$ & $\log M_{\rm BH,dyn}$ & $\lambda_{\rm int}$ \\ 
(1) & (2) & (3) & (4) & (5) & (6) & (7) & (8) & (9) & (10) & (11) & (12) & (13) & (14) & (15) \\
\midrule
\endhead

\midrule
\multicolumn{14}{r}{{Continued on next page}} \\
\endfoot

\bottomrule
\endlastfoot

0013$+$5535 & 00 15 40.13 & $+$55 51 44.7 & 0.109 & 15.76 & $3.05^*$ &  & 1 & 43.406 & 42.708 & 41.030 & 40.062 & 0.719 & 9.68 & $-5.086$\\
0032$+$595 & 00 35 52.63 & $+$59 50 04.3 & 0.086 & 17.05 & 3.19 & 2.18 & 2 & 44.308 & 43.609 & 40.609 & 39.601 & 0.563 & $7.25$ & $-1.754$\\
0113$+$2504 & 01 15 46.15 & $+$25 19 53.4 & 0.376 & 15.75 & 2.59 &  & 3 & 45.386 & 44.688 & 41.826 & 40.999 & 0.579 & 9.03 & $-2.456$\\
0120$+$340 & 01 23 08.63 & $+$34 20 48.4 & 0.272 & 17.64 & 4.75 & 2.05 & 4 & 45.549 & 44.851 & 41.558 & 40.204 & 0.528 & 8.68 & $-1.943$\\
0145$+$138 & 01 48 29.71 & $+$14 02 17.9 & 0.125 & 15.49 & 1.72 & 1.99 & 5 & 43.393 & 42.700 & 39.615 & 39.144 & 0.554 & 8.42 & $-3.834$\\
0150$+$0132 & 01 52 39.61 & $+$01 47 17.3 & 0.080 & 16.58 & $3.33^*$ & 2.82 & 6 & 43.631 & 42.932 & 40.712 & 39.667 & 0.655 & 9.34 & $-4.522$\\
0156$+$1032 & 01 59 34.39 & $+$10 47 05.7 & 0.195 & 15.80 & $3.22^*$ &    & 3 & 43.992 & 43.293 & 41.270 & 40.254 & 0.678 & 8.51 & $-3.331$\\
0158$+$003 & 02 01 06.18 & $+$00 33 59.9 & 0.299 & 16.91 & 3.71 & 2.28 & 5 & 44.877 & 44.179 & 41.116 & 39.977 & 0.555 & 8.19 & $-2.125$\\
0208$+$352 & 02 08 38.16 & $+$35 23 12.7 & 0.318 & 16.34 & 2.75 &  & 7 & 44.683 & 43.985 & 40.923 & 40.044 & 0.556 & 8.62 & $-2.749$\\
0210$+$515 & 02 14 17.93 & $+$51 44 51.9 & 0.049 & 17.18 & 1.50 & 2.08 & 4 & 43.754 & 43.063 & 40.660 & 40.308 & 0.634 & 8.55 & $-3.601$\\
0227$+$020 & 02 27 16.58 & $+$02 01 59.9 & 0.457 & 15.44 & 2.41 & 1.87 & 8 & 45.491 & 44.795 & 41.551 & 40.787 & 0.534 & 9.47 & $-2.789$\\
0229$+$200 & 02 32 48.61 & $+$20 17 17.4 & 0.139 & 19.05 & 1.93 & 1.84 & 9 & 44.669 & 43.976 & 41.067 & 40.496 & 0.574 & 10.08 & $-4.218$\\
0301$-$243 & 03 03 26.50 & $-$24 07 11.4 & 0.266 & 15.40 & 5.44 &  & 10 & 45.171 & 44.472 & 42.402 & 40.931 & 0.673 & 8.95 & $-2.592$\\
0317$+$185 & 03 19 51.81 & $+$18 45 34.5 & 0.190 & 16.91 & 2.37 & 2.00 & 5 & 44.860 & 44.163 & 40.944 & 40.194 & 0.537 & 8.66 & $-2.611$\\
0323$+$022 & 03 26 13.94 & $+$02 25 14.6 & 0.147 & 15.90 & 2.07 & 2.54 & 11 & 44.069 & 43.371 & 40.768 & 40.136 & 0.610 & 9.21 & $-3.953$\\
0347$-$121 & 03 49 23.18 & $-$11 59 27.2 & 0.188 & 18.26 & $3.75^*$ & 2.03 & 5 & 44.797 & 44.098 & 40.627 & 39.479 & 0.507 & 8.60 & $-2.616$\\
0350$-$371 & 03 51 54.54 & $-$37 03 44.3 & 0.165 & 17.54 & 2.96 &  & 1 & 44.261 & 43.563 & 40.723 & 39.781 & 0.582 & 8.82 & $-3.371$\\
0414$+$009 & 04 16 52.49 & $+$01 05 23.8 & 0.287 & 16.69 & 2.16 & 2.49 & 12 & 45.023 & 44.325 & 41.583 & 40.914 & 0.593 & 8.40 & $-2.189$\\
0502$+$675 & 05 07 56.25 & $+$67 37 24.4 & 0.314 & 19.18 & 3.55 & 2.43 & 4 & 45.888 & 45.190 & 41.729 & 40.629 & 0.509 & 9.00 & $-1.924$\\
0506$-$039 & 05 09 38.18 & $-$04 00 45.8 & 0.304 & 17.18 & 1.93 & 2.14 & 5 & 45.220 & 44.524 & 41.641 & 41.069 & 0.577 & 8.75 & $-2.340$\\
0548$-$322 & 05 50 40.56 & $-$32 16 16.3 & 0.069 & 17.36 & 2.10 & 2.01 & 13 & 44.460 & 43.764 & 40.667 & 40.022 & 0.552 & 8.74 & $-3.090$\\
0645$+$1520 & 06 48 47.65 & $+$15 16 24.8 & 0.179 & 16.89 & $4.97^*$ & 2.51 & 14 & 44.458 & 43.759 & 41.412 & 40.051 & 0.640 & 8.84 & $-3.195$\\
0647$+$250 & 06 50 46.49 & $+$25 02 59.6 & 0.203 & 16.30 & 3.24 & 2.15 & 15 & 45.143 & 44.445 & 41.617 & 40.596 & 0.583 & 7.73 & $-1.399$\\
0706$+$592 & 07 10 30.96 & $+$59 08 20.4 & 0.125 & 17.72 & 1.87 &  & 16 & 43.922 & 43.225 & 40.845 & 40.301 & 0.636 & 9.75 & $-4.639$\\
0737$+$746 & 07 44 05.26 & $+$74 33 57.6 & 0.315 & 16.50 & 4.95 & 2.28 & 17 & 44.431 & 43.732 & 41.536 & 40.147 & 0.658 & 9.94 & $-4.322$\\
0806$+$524& 08 09 49.18 & $+$52 18 58.2 & 0.138 & 15.80 & 2.51 &  & 18 & 43.926 & 43.288 & 41.227 & 40.427 & 0.681 & 8.90 & $-3.786$\\
0906$+$313 & 09 09 53.28 & $+$31 06 03.1 & 0.272 & 16.68 & $4.29^*$ &  & 3 & 44.949 & 44.250 & 41.959 & 40.694 & 0.647 & 8.91 & $-2.774$\\
0927$+$500 & 09 30 37.57 & $+$49 50 25.5 & 0.187 & 17.30 & 2.61 & 2.11 & 5 & 44.819 & 44.121 & 40.874 & 40.041 & 0.534 & 8.87 & $-2.863$\\
1011$+$496 & 10 15 04.14 & $+$49 26 00.7 & 0.212 & 15.60 & 2.80 & 2.30 & 19 & 45.460 & 44.762 & 41.723 & 40.829 & 0.558 & 8.94 & $-2.292$\\
1028$+$511 & 10 31 18.52 & $+$50 53 35.8 & 0.360 & 16.77 & 3.35 & 2.11 & 12 & 45.605 & 44.907 & 41.712 & 40.662 & 0.540 & 8.07 & $-1.277$\\
1038$+$392 & 10 41 49.14 & $+$39 01 19.5 & 0.208 & 16.68 & $2.58^*$ &  & 3 & 44.531 & 43.833 & 41.163 & 40.340 & 0.602 & 8.73 & $-3.011$\\
1050$+$4946 & 10 53 44.12 & $+$49 29 55.9 & 0.140 & 15.80 & $2.27^*$ &  & 3 & 43.602 & 42.905 & 40.883 & 40.171 & 0.679 & 8.74 & $-3.949$\\
1101$+$384 & 11 04 27.31 & $+$38 12 31.7 & 0.030 & 16.22 & 2.80 &  & 16 & 44.092 & 43.393 & 40.731 & 39.836 & 0.603 & 9.05 & $-3.771$\\
1133$+$6753 & 11 36 30.08 & $+$67 37 04.3 & 0.134 & 17.48 & 2.64 & 2.04 & 4 & 44.856 & 44.159 & 40.981 & 40.138 & 0.542 & 8.67 & $-2.625$\\
1133$+$704 & 11 36 26.40 & $+$70 09 27.3 & 0.045 & 15.76 & 1.84 &  & 16 & 43.833 & 43.137 & 40.494 & 39.964 & 0.605 & 8.95 & $-3.927$\\
1219$+$305 & 12 21 21.94 & $+$30 10 37.1 & 0.184 & 16.27 & 2.61 & 2.44 & 13 & 45.402 & 44.703 & 41.430 & 40.597 & 0.531 & 9.00 & $-2.411$\\
1221$+$2452 & 12 24 24.18 & $+$24 36 23.4 & 0.219 & 15.55 & 2.96 &  & 10 & 44.848 & 44.149 & 41.174 & 40.232 & 0.566 & 8.29 & $-2.255$\\
1229$+$645 & 12 31 31.39 & $+$64 14 18.2 & 0.163 & 15.84 & 4.08 &  & 20 & 44.300 & 43.601 & 41.194 & 39.973 & 0.633 & 8.96 & $-3.473$\\
1235$+$632 & 12 37 39.07 & $+$62 58 42.8 & 0.297 & 16.24 & 3.02 & 1.92 & 8 & 44.531 & 43.834 & 41.301 & 40.341 & 0.618 & 8.64 & $-2.920$\\
1255$+$244 & 12 57 31.93 & $+$24 12 40.2 & 0.141 & 18.68 & 1.30 & 2.04 & 8 & 44.438 & 43.751 & 40.272 & 40.045 & 0.508 & 8.56 & $-2.923$\\
1415$+$259 & 14 17 56.67 & $+$25 43 26.0 & 0.240 & 17.06 & 2.13 & 2.11 & 4 & 45.354 & 44.657 & 41.544 & 40.887 & 0.550 & 8.17 & $-1.627$\\
1421$+$582 & 14 22 38.86 & $+$58 01 55.5 & 0.683 & 17.48 & 1.51 &  & 3 & 46.077 & 45.382 & 41.800 & 41.442 & 0.495 & 9.20 & $-1.932$\\
1426$+$428 & 14 28 32.60 & $+$42 40 20.9 & 0.129 & 18.01 & 1.56 & 1.92 & 13 & 45.107 & 44.417 & 40.686 & 40.299 & 0.478 & 8.59 & $-2.287$\\
1437$+$397 & 14 39 17.46 & $+$39 32 42.8 & 0.344 & 15.86 & 3.31 &  & 3 & 44.945 & 44.247 & 41.883 & 40.844 & 0.638 & 8.79 & $-2.657$\\
1440$+$122 & 14 42 48.21 & $+$12 00 40.2 & 0.163 & 16.90 & 2.06 &  & 10 & 44.652 & 43.954 & 41.178 & 40.551 & 0.590 & 8.74 & $-2.900$\\
1517$+$656 & 15 17 47.58 & $+$65 25 23.2 & 0.702 & 16.63 & 3.31 & 2.60 & 5 & 46.343 & 45.644 & 42.330 & 41.290 & 0.526 & 10.11 & $-2.580$\\
1518$+$407 & 15 18 38.89 & $+$40 45 00.2 & 0.065 & 15.70 & $2.06^*$ &   & 3 & 42.774 & 42.076 & 40.172 & 39.544 & 0.693 & 8.25 & $-4.288$\\
1533$+$189 & 15 33 11.25 & $+$18 54 29.0 & 0.307 & 17.20 & $2.96^*$ &  & 3 & 44.904 & 44.205 & 41.418 & 40.476 & 0.588 & 8.91 & $-2.819$\\
1533$+$535 & 15 35 00.79 & $+$53 20 37.3 & $0.560^\dagger$ & 16.94 & 2.55 &  & 16 & 45.997 & 45.299& 41.810 & 40.997 & 0.505 & 10.21 & $-3.025$\\
1544$+$820 & 15 40 15.87 & $+$81 55 05.7 & $0.460^\dagger$ & 15.98 & 2.30 &  & 16 & 46.190 & 45.492 & 42.237 & 41.513 & 0.533 & 7.82 & $-0.442$\\
1552$+$203 & 15 54 24.13 & $+$20 11 25.4 & 0.222 & 17.40 & 2.06 &  & 21 & 44.628 & 43.931 & 41.370 & 40.743 & 0.615 & 8.94 & $-3.123$\\
1624$+$3520 & 16 26 25.85 & $+$35 13 41.4 & 0.498 & 15.74 & 2.53 & 2.50 & 8 & 44.580 & 43.882 & 41.832 & 41.026 & 0.675 & 8.82 & $-3.052$\\
1652$+$398 & 16 53 52.21 & $+$39 45 36.6 & 0.033 & 15.45 & 4.72 & 2.23 & 20 & 44.052 & 43.353 & 41.238 & 39.890 & 0.668 & 9.91 & $-4.671$\\
1722$+$119 & 17 25 04.34 & $+$11 52 15.4 & 0.180 & 15.40 & 1.13 &  & 16 & 45.105 & 44.418 & 41.606 & 41.499 & 0.586 & 9.87 & $-3.566$\\
1727$+$502 & 17 28 18.62 & $+$50 13 10.4 & 0.055 & 16.34 & 1.91 & 2.20 & 22 & 43.651 & 42.955 & 40.685 & 40.123 & 0.649 & 8.28 & $-3.439$\\
1741$+$196 & 17 43 57.83 & $+$19 35 09.0 & 0.084 & 17.24 & 3.19 &  & 16 & 44.530 & 43.832 & 41.266 & 40.258 & 0.614 & 9.69 & $-3.972$\\
1757$+$703 & 17 57 13.07 & $+$70 33 37.8 & 0.407 & 17.30 & 3.02 &  & 23 & 45.105 & 44.407 & 41.517 & 40.557 & 0.576 & 8.75 & $-2.457$\\
1837$+$4759 & 18 38 49.15 & $+$48 02 34.3 & 0.300 & 15.76 & 2.45 &  & 1 & 44.800 & 44.102 & 41.527 & 40.748 & 0.613 & 8.22 & $-2.232$\\
1959$+$650 & 19 59 59.85 & $+$65 08 54.6 & 0.047 & 15.96 & 5.20 & 2.29 & 19 & 44.071 & 43.372 & 40.817 & 39.385 & 0.616 & 9.07 & $-3.812$\\
2037$+$521 & 20 39 23.51 & $+$52 19 50.1 & 0.053 & 16.76 & 3.55 &  & 1 & 42.823 & 42.124 & 40.029 & 38.929 & 0.670 & 8.65 & $-4.640$\\
2039$+$2416 & 20 42 06.04 & $+$24 26 52.3 & 0.104 & 17.18 & $2.34^*$ &  & 16 & 44.334 & 43.637 & 40.765 & 40.026 & 0.578 & 8.76 & $-3.237$\\
2143$+$070 & 21 45 52.30 & $+$07 19 27.2 & 0.237 & 17.50 & 2.53 &  & 1 & 44.540 & 43.842 & 41.583 & 40.777 & 0.651 & 8.46 & $-2.732$\\
2155$-$304 & 21 58 52.06 & $-$30 13 32.1 & 0.117 & 15.75 & 3.34 & 2.39 & 24 & 44.914 & 44.215 & 41.795 & 40.748 & 0.631 & 8.91 & $-2.808$\\
2247$+$381 & 22 50 05.74 & $+$38 24 37.1 & 0.119 & 16.36 & 2.03 &  & 10 & 44.095 & 43.398 & 41.045 & 40.431 & 0.640 & 9.44 & $-4.156$\\
2313$+$147 & 23 13 57.34 & $+$14 44 23.3 & 0.164 & 17.07 & 4.42 &  & 3 & 43.865 & 43.167 & 40.986 & 39.696 & 0.660 & 9.11 & $-4.057$\\
2316$-$432 & 23 19 05.89 & $-$42 06 48.3 & 0.055 & 15.56 & $1.76^*$ &  & 25 & 42.803 & 42.107 & 40.704 & 40.212 & 0.752 & 9.41 & $-5.417$\\
2320$+$343 & 23 22 44.02 & $+$34 36 13.8 & 0.098 & 16.03 & 1.23 &  & 3 & 43.056 & 42.367 & 40.563 & 40.384 & 0.705 & 9.55 & $-5.297$\\
2344$+$514 & 23 47 04.83 & $+$51 42 17.8 & 0.044 & 16.19 & 3.63 & 2.28 & 4 & 43.734 & 43.035 & 40.683 & 39.563 & 0.640 & 7.71 & $-2.789$\\
2356$-$309 & 23 59 07.90 & $-$30 37 40.6 & 0.165 & 17.29 & $2.58^*$ & 2.20 & 26 & 44.840 & 44.142 & 41.040 & 40.216 & 0.551 & 8.60 & $-2.572$\\
\end{longtable}
\vspace{0.5cm}
\footnotesize
\begin{minipage}{\linewidth}
Col.\,(1) IAU Name; Col.\,(2) and Col.\,(3) are Right Ascension and Declination (J2000); Col.\,(4) Redshift ("$\dagger$" denotes that the HBLs have featureless optical spectra; consequently, their redshift is given in the literature only as an upper or lower limit \citep[see][]{Sbarufatti2005ApJ,Sheng2024ApJ}); Col.\,(5) Logarithm of the synchrotron-peak frequency for HBLs (Hz); Col.\,(6) The 5\,GHz Doppler factor ("*" denotes that the 5\,GHz Doppler factors are estimated by us); Col.\,(7) The power law photon index (For the HBLs without $\Gamma_{\rm{X}}$, we adopted the mean $\Gamma_{\rm{X}}=2.34$ for HBLs from \cite{Donato2001hard}); Col.\,(8) Reference of $\Gamma$ and the observational X-ray flux ($F_{\rm{X,obs}}$); Col.\,(9) Logarithm of the observational 2--10\,keV X-ray luminosity ($\rm erg\,s^{-1}$); Col.\,(10) Logarithm of the intrinsic 2--10\,keV X-ray luminosity that are corrected from $L_{\rm{X,obs}}$ using the Doppler factor ($\rm erg\,s^{-1}$); Col.\,(11) Logarithm of the observational 5\,GHz core radio luminosity ($\rm erg\,s^{-1}$); Col.\,(12) Logarithm of the intrinsic 5\,GHz core radio luminosity that are corrected from $L_{\rm{R,obs}}$ using the Doppler factor  ($\rm erg\,s^{-1}$); Col.\,(13) The broad-band spectral index from 5\,GHz radio band to 2--10\,keV X-ray band; Col.\,(14) Logarithm of dynamic BH mass ($M_\odot$); Col.\,(15) Eddington-ratio: $\lambda_{\rm{int}}=\log(L_{\rm{X,int}}/L_{\rm{Edd}})$.\\
References: (1) \cite{Saxton2008first}; (2) \cite{Gianni2011x}; (3) \cite{Dwelly2017spiders}; (4) \cite{Ricci2017bat}; (5) \cite{Donato2005six}; (6) \cite{Aharonian2008discovery}; (7) \cite{Ueda2005asca}; (8) \cite{Massaro2011x}; (9) \cite{Kaufmann2011-1ES}; (10) \cite{Wierzcholska2016x}; (11) \cite{Perlman2005}; (12) \cite{Fang2005high}; (13) \cite{Blustin2004}; (14) \cite{Aliu2011}; (15) \cite{Acciari2023long}; (16) \cite{Warwick2012xmm}; (17) \cite{Piconcelli2002}; (18) NED; (19) \cite{Giommi2021x}; (20) \cite{Gonzalez2012x}; (21) \cite{Marchesi2025}; (22) \cite{Aleksic2014}; (23) \cite{Sazonov2024}; (24) \cite{Foschini2008}; (25) \cite{Ueda2001asca}; (26) \cite{Abramowski2010multi}.
\end{minipage}
\end{landscape}

\clearpage
\twocolumn


\section*{Data Availability}

For this work, the available core 5\,GHz flux density, X-ray flux and dynamic BH mass of HBLs are taken from NED and referenced literatures (see \S\,\ref{Sample}). The available data underlying this article are provided in table\,\ref{table2}.

\bibliographystyle{mnras}
\bibliography{example} 




\appendix


\bsp	
\label{lastpage}
\end{document}